\documentclass[aps,pra,showpacs,twoside,twocolumn,10pt]{revtex4-1}
\usepackage[colorlinks=true, citecolor=red, urlcolor=blue ]{hyperref}
\usepackage{epsfig,newlfont,amssymb,amsfonts,amsmath,bm,subfigure,palatino,mathtools,amsthm,braket,times,soul}
\usepackage{color}
\usepackage{multirow}
\usepackage{ulem}
\usepackage{physics}
\usepackage{mathrsfs}
\usepackage{graphicx}
\usepackage{mathtools}
\usepackage{ulem}
\usepackage{cancel}

\definecolor{indiagreen}{rgb}{0.07, 0.53, 0.03}
	\definecolor{teal}{rgb}{0.0, 0.53, 0.53}

\begin{document}

\title{Statistics of Entanglement Transformation with Hierarchies among  Catalysts}

	\author{Rivu Gupta\(^1\), Arghya Maity\(^1\), Shiladitya Mal\(^{2,3}\), and Aditi Sen(De)\(^{1}\)}
	
		\affiliation{$^1$ Harish-Chandra Research Institute and HBNI, Chhatnag Road, Jhunsi, Allahabad - 211019, India\\
	$^2$ Physics Division, National Center for Theoretical Sciences, Taipei 10617, Taiwan \\
$^3$ Department of Physics and Center for Quantum Frontiers of Research and Technology (QFort),
National Cheng Kung University, Tainan 701, Taiwan	}

\begin{abstract}

The distribution of typical bipartite pure states  is studied within the framework of state transformation via local operations and classical communication (LOCC). We report the statistics of comparable and incomparable states in different dimensions for single- and multi-copy regimes and we establish a connection between  state transformation and the difference between the entanglement contents  of the initial and the target states. From the analysis of catalyst resources, required to 
 further otherwise impossible LOCC transformations between pairs, we demonstrate  a universal pattern in the average and minimum entanglement of the randomly generated catalysts. 
 Furthermore, we  introduce a concept of hierarchy between different kinds of catalysts, and we show how they can not only aid in the conversion of  incomparable states, but they can also act as a less costly resource toward this goal. We confirm the existence of catalysts, referred to as \textit{strong  catalysts},  which can activate LOCC transformation between pairs at the single-copy level, when it is initially impossible even  with  multiple copies.

\end{abstract}

	
\maketitle
	
\section{Introduction}
\label{intro}
Quantum entanglement is one of the most functional resources in quantum information tasks ranging from quantum communication \cite{Bennett_PRL_1992, Bennett_PRL_1993, Ekert_PRL_1991}, genuine randomness certification \cite{Pironio_Nature_2010}, to  one-way quantum computing \cite{Raussendorf_PRL_2001, Briegel_Nature_2009}. 
Therefore, it is important to develop a procedure to manipulate entanglement in quantum states, so that they may be brought into a useful form to fulfill a particular task.  
Entanglement transformation plays a crucial role in the manipulation process, by which a potentially less applicable state is turned into a more useful one, by  some non-resource generating operations, say, local operations and classical communication (LOCC). 

In the LOCC state conversion domain, three types of state transformation are typically studied --  deterministic \cite{Nielsen_PRL_1999}, probabilistic \cite{Jonathan_PRL_1999_B} and approximate state transformations  \cite{Jonathan_PRA_2000}. Deterministic transformation, in which a pure state is mapped to another one with unit probability under LOCC, is dictated by the celebrated necessary and sufficient criterion by Neilsen based on majorization \cite{CoverThomasbook} (for multipartite regime, 
see \cite{Xin_PRA_2007}). Further, Jonathan and Plenio \cite{Jonathan_PRL_1999_B} introduced probabilistic entanglement transformation  and the corresponding maximum probability for bipartite pairs can also be obtained \cite{Vidal_PRL_1999} (cf.  \cite{Kang_PRA_2021}). By using majorization criteria,  several conditions are derived for non-asymptotic state transformation involving multiple copies of the initial and final states \cite{Ujjwal_PRA_2002} while asymptotic state conversion has also been extensively studied in the state transformation protocol of quantum resource theory  \cite{Bennett_PRA_1996, Peter_JMP_2015, Streltsov_PRL_2020}. On the other hand,  approximate transformation \cite{Jonathan_PRA_2000, Chandan_arxiv_2021} 
identifies the best possible final state which is nearest in terms of the fidelity to the  desired target state. 

 When transformation between a given pair of states is not possible via LOCC, an additional entangled state known as the \textit{catalyst}, is introduced to assist the protocol, which  remains unaltered after the transformation just like in a chemical process
\cite{Jonathan_PRL_1999_A}. Catalysts are also shown to develop correlations with the states and  enhance the transformation probability and  power \cite{Chandan_PRL_2021}. 
The  role of catalysts has also been investigated in the context of quantum thermodynamics  \cite{Jonathan_Oppenheim_NAS_2015, Lipka_PRX_2021},  quantum coherence \cite{Bu_PRA_2016, Streltsov_RMP_2017, Aberg_PRL_2014}, purity \cite{Gour_PR_2015}, and state symmetries \cite{Marvian_NJP_2013}. A few attempts have been made to characterize the general properties of catalysts which include  
a lower bound for the dimension of the catalysts \cite{Gilad_PRA_2009}  as well as the maximum and minimum bounds on entanglement of the catalysts \cite{Gilad_PRA_2019}. 
Another kind of catalysis in which the catalysts are not returned in their original form, is called inexact catalysis. A set of such optimal inexact catalysts with minimal trace distance error that can be used in a state transformation process is also provided \cite{Ng_NJP_2015}.

On the other hand, random states play an important role in quantum information science which can arise naturally when the system interacts with the environment. It was shown that instead of random behavior, the features like entanglement entropy, purity, multipartite entanglement, classical correlations of random states \cite{Kendon_PRA_2002, Patrick_CMP_2006,  Hastings_NaturePhysics_2009, Gross_PRL_2009, Winter_PRL_2009, Marco_Entropy_2018, Waldemar_PRA_2019,  Aditi_PRA_2019, Aditi_PRA_2020} can follow a pattern  \cite{bengtsson_zyczkowski_2006}. In particular, it was demonstrated that most of the typical states  become highly multiparty entangled with an increase of the number of parties. 
 Moreover, these states have also been  used  to disprove the additivity of minimal output entropy \cite{Benjamin_PRA_1996}, to obtain the maximal purity in \(k\)-uniform  states with \(N\) parties \cite{Waldemar_PRA_2019},  in showing constructive feedback in presence of a non-Markovian noisy environment \cite{GuptaCF'21}, putting restriction on classical correlation \cite{Roy'21}, and analysing power of teleportation and densecodability of bipartite channels \cite{Gupta'21}.  
 For our purpose, we are interested in bipartite pure states of arbitrary dimensions chosen at random with respect to the uniform measure, called Haar measure, on the unit sphere in a finite-dimensional Hilbert space that is invariant under all unitary transformations. 

 In this work, we characterize the pattern in entanglement transformation obtained from typical pairs of states and catalysts.
 Specifically, we study the distributions of comparable as well as incomparable bipartite pure states simulated randomly in different dimensions, and the corresponding behavior with the increasing number of copies. We prove that the LOCC transformability is related to the entanglement difference between the source and the target states which can be thought of as the resource consumed in the transformation protocol. For a given incomparable pair, we introduce three quantities to analyze the entanglement properties of the catalyst states found from the generated random states -- the mean and the minimum entanglement of the catalyst as well as the distribution in the average number of catalyst states. We observe that the average entanglement required to catalyze a pair of incomparable states decreases with the increase in the difference of the entanglement content for the input-output pair while it increases with the dimension. Moreover,  when a higher number of copies of the input-output pairs are available, the incomparable pair with higher number of copies requires catalyst with low entanglement content compared to that of the single-copy incomparable ones.


Comparing multiple-copy state transformation and the role of catalysts in the transformation via LOCC, we introduce the concept of hierarchy between various kinds of catalyst states. Specifically, we try to understand the characteristics of catalysts which can assist state transformations with different levels of difficulty. We introduce the notion of \textit{strong catalysts}, which can promote certain transformations at the single-copy level, even when it is originally impossible in the presence of multiple pairs of the source and target states. Such states are especially resourceful in the sense that  they can assist transformations of pairs which remain incomparable even  when large amounts of entanglement is available in the form of multiple copies.
We aim to determine the entanglement properties of such \textit{strong catalysts}  and investigate the distribution of the average number of such catalysts obtained from randomly generated states with respect to the entanglement of the pairs to be transformed. 
The entire investigation 
sheds light on the transformation capabilities of states when they are sampled at random from the state space and establish a uniform pattern on the resource available that can dictate the LOCC transformability . 

 The rest of the paper is organised in the following way. In Sec. \ref{sec:pre}, we lay down the basic ideas required to explain our results which include the Haar-uniform generation of random states and the theory of state transformation followed by catalyst assisted state transformation protocol. We then analyse the behavior of entanglement for randomly generated pure states  by varying the dimension  in Sec. \ref{sec:ent}. The following section, Sec. \ref{sec:1}, deals with the transformation properties of random bipartite states through LOCC in different dimensions and elucidate the relationship with the entanglement required to facilitate such transformations. Sec. \ref{sec:2} shows how random states can act as catalysts for state transformations in different dimensions. We analyse the entanglement properties of such catalyst states in the single- and the multi-copy regimes in Subsecs. \ref{subsec:singlecopy} and \ref{subsec:multiplecopy} respectively. We compare catalysts for various types of state transformation routines, thereby establishing a hierarchy between such states in terms of their catalytic abilities in Sec. \ref{sec:cat_hie}. We end our paper with conclusions in Sec. \ref{sec:conclu}.

\section{Prerequisites}
\label{sec:pre}

In this section,  we first discuss the Haar uniform generation of   pure two-qudit states and  briefly describe the criteria for deterministic state transformation under local operations and classical communication. We then shift our attention to the paradigm of  transformation between the initial and final states with the help of catalysts, recounting the necessary conditions to implement such protocols.

\subsection{Generation of random states}
\label{subsec:haar}

Given a basis, quantum states  are  specified by complex coefficients. To generate states Haar uniformly in the state space \cite{bengtsson_zyczkowski_2006}, we randomly simulate real  numbers from a Gaussian distribution with vanishing mean  and unit standard deviation, denoted as \(G(0,1)\).
	In particular, a two-qudit pure state in $\mathbb{C}^d \otimes \mathbb{C}^d$ (denoted in short as \(d \otimes d\)) can be represented as 
	\begin{equation}
	|\psi \rangle_{AB} = \sum_{i,j=1}^{d} (a_{ij} + i b_{ij}) |i \rangle_{A} \otimes |j \rangle_{B},
	\label{eq:pure_random}
	\end{equation}
	where $a_{ij}$ and $b_{ij}$ are real numbers sampled from \(G(0,1)\) \cite{Zyczkowski_JMP_2011,bengtsson_zyczkowski_2006} and $d$ denotes the dimension of the system. \(\{|i\rangle\}\) (\(\{|j\rangle\}\)) denotes the computational basis for the first (second) party. The parameters $a_{ij}$ and $b_{ij}$ thus belong to the distributions, given by
\(\Phi(a_{ij}) = \frac{1}{\sqrt{2\pi}} \exp (-\frac{a^{2}_{ij}}{2})\), and \(\Phi(b_{ij}) = \frac{1}{\sqrt{2\pi}} \exp (-\frac{b^{2}_{ij}}{2})\).
	Such a generation ensures that the states selected are not confined to a particular region of the state space.
The coefficients form a $2d^2$ variable tuple which we can depict as $r = \{a_{ij}, b_{ij}\}$, $r \in \mathbb{R}^{d^2}$, the \(d^2\)-dimensional real space. The joint probability distribution for $r$ thus reads as
\(f(r) = (\frac{1}{2\pi})^{d^2/2} \exp [ - \sum_{i,j=1}^{d}  (a^{2}_{ij} + b^{2}_{ij})/2  ] 
 = (\frac{1}{2\pi})^{d^2/2} \exp [ - |r|^2 /2  ]\) 
which implies that $f(r)$ depends only on the length of $r$ and not its direction. This means that $\hat{r} = r/|r|$ is distributed uniformly over the \(d^2-1\)-dimensional spherical surface in $\mathbb{R}^{d^2}$. This further implies that $f( \hat{r})$ remains constant in all directions over this subspace i.e., the states are generated Haar uniformly
(see 
\cite{Dahlsten_JPA_2014} and \cite{Ozols_essay_2009}).


	\subsection{ Deterministic transformation between pure states  via LOCC}
	\label{subsec:state_transf}

	The aim of  deterministic state transformation is to convert an entangled state shared between two spatially separated parties to another entangled state by using local operations and classical communication.  When a state, $|\psi\rangle_{AB}$ transforms to another state, $|\phi\rangle_{AB}$ with certainty, $E(|\psi\rangle_{AB}) \geq E(|\phi\rangle_{AB})$, where $E$ is an appropriate measure of entanglement for pure states \cite{Bennett_PRA_1996} since entanglement cannot be increased by LOCC.  In a seminal paper by Nielsen  \cite{Nielsen_PRL_1999}, it was shown that such deterministic state transformation 
	can occur according to the majorization condition
	 for bipartite pure states.
	 
	 If $\boldsymbol{i}$ and $\boldsymbol{j}$ are two vectors in  $\mathbb{R}^{\otimes d}$, $\boldsymbol{i}$ is majorized by $\boldsymbol{j}$, denoted by $\boldsymbol{i \prec j}$, if they satisfy the condition, $\sum_{l=1}^{n} i_l^{\downarrow} \leq \sum_{l=1}^{n} j_l^{\downarrow} ~ \forall n \in { 1,2,3,....d}$, where $i_l^{\downarrow}$ and $j_l^{\downarrow}$ indicate that the elements are ordered in a non-increasing manner \cite{CoverThomasbook}. We can write any two pure bipartite states in their Schmidt form as $ \ket{\psi}_{AB} =\sum_{l=1}^{d} \sqrt{\alpha_l}\ket{\gamma}_A\ket{\gamma}_B$ and $ \ket{\phi}_{AB}=\sum_{k=1}^{d} \sqrt{\beta_k}\ket{\eta}_A\ket{\eta}_B$, where the Schmidt coefficients $\alpha_l$ and $\beta_k$ satisfy the properties $\alpha_l, \beta_k \geq 0$ and $\sum_{l=1}^{d} \alpha_l = 1 $ and  $\sum_{k=1}^{d} \beta_k = 1 $. We can now define two sets of ordered Schmidt coefficients (OSCs) as $\boldsymbol{\alpha} = \{\alpha_i^\downarrow\}$ and $\boldsymbol{\beta} = \{\beta_i^\downarrow\}$. According to Nielsen's theorem, the initial state $\ket{\psi}_{AB}$ can be transformed to the final state $\ket{\phi}_{AB}$ via LOCC, i.e.,   $\ket{\psi}_{AB}$ $\xrightarrow[LOCC]{}$ $\ket{\phi}_{AB}$ if and only if $\boldsymbol{\alpha \prec \beta}$, thereby providing a necessary and sufficient condition for $\ket{\psi}_{AB}$  to be converted into $\ket{\phi}_{AB}$  with unit probability. Depending on whether a pair of states can be converted to one another by LOCC, they are called \textit{comparable} ($\ket{\psi}_{AB}\xrightarrow[LOCC]{} \ket{\phi}_{AB}$) or \textit{incomparable} ($\ket{\psi}_{AB} \nleftrightarrow_{LOCC} \ket{\phi}_{AB} $) states respectively. 

	 Instead of deterministic transformation, one can consider probabilistic state transformations, i.e., although
	 the initial and the final states  are incomparable deterministically,  their conversion is possible by LOCC with a non-zero finite probability  \cite{Vidal_PRL_1999, Jonathan_PRL_1999_B}. 
	On the other hand, if a pair of states are not convertible at the single-copy level, they can  be comparable when multiple copies are considered \cite{Ujjwal_PRA_2002}, 
	 i.e., although $\ket{\psi}_{AB} \not\to_{LOCC} \ket{\phi}_{AB}$,  $\ket{\psi}_{AB}^{\otimes k} \xrightarrow[LOCC]{} \ket{\phi}_{AB}^{\otimes k}$, where $k$ denotes the number of copies considered for each state. The additional copies of the states make sure that the eventual OSCs $\boldsymbol{\alpha^{\otimes k}}$ and $\boldsymbol{\beta^{\otimes k}}$ satisfy the majorization condition. In our work, we mainly conduct our analysis within the regime of deterministic transformations governed by Nielsen's majorization criterion for single as well as multiple copies of the initial and final states.
	
	\subsection{Catalytic entanglement transformation}
	\label{subsec:cata_transf}

	The concept of catalysis was introduced by Jonathan 
	and Plenio \cite{Jonathan_PRL_1999_A} 
	in entanglement state transformation. If $|\psi\rangle_{AB} \not\to_{LOCC} |\phi\rangle_{AB}$, i.e., $\boldsymbol{\alpha} \nprec \boldsymbol{\beta}$,  an additional entangled state $|\mathbf{\chi}\rangle_{AB} = \sum_{i =1}^{d} \sqrt{c_i} |\zeta\rangle_A |\zeta\rangle_B$  having OSCs $\boldsymbol{\chi} = \{c_i^\downarrow\}$ can be introduced so that $|\psi\rangle_{AB} \otimes |\chi\rangle_{AB} $ can be mapped to $|\phi\rangle_{AB} \otimes |\chi\rangle_{AB}$ by LOCC $\Rightarrow \boldsymbol{\alpha} \otimes \boldsymbol{\chi} \prec \boldsymbol{\beta} \otimes \boldsymbol{\chi}  $. Here \(|\chi\rangle_{AB}\) is called the catalyst. 
	Notice that the catalyst state remains unchanged at the end of the protocol, thereby highlighting the use of entanglement as a resource without consuming it in any way. It can be shown that maximally entangled states in $d \otimes d$ having the form $|\chi\rangle_{AB} = \frac{1}{\sqrt{d}} \sum_{i = 1}^d |i\rangle_A |i\rangle_B$ can never act as a successful catalysts. Furthermore, two bipartite states in $d$-dimensions, characterised by OSCs $\{\alpha_i^\downarrow\}~\text{and}~\{\beta_i^\downarrow\}$ (\(i=1,2, \ldots, d\)) are convertible through catalyst assistance if and only if $\alpha_1 \leq \beta_1$ and $\alpha_d \geq \beta_d$ \cite{Jonathan_PRL_1999_A}. From this condition, it immediately implies that  for a pair of incomparable two-qutrit states, catalysts do not exist. It was also demonstrated that catalytic assistance can be helpful to increase the maximum probability of transformation between the initial and the final states, even if the conversion protocol cannot be deterministically implemented.

\section{
Entanglement distribution for random Qudits }
\label{sec:ent}
\begin{figure}[h]
		\centering
		\includegraphics[width=1.05\linewidth]{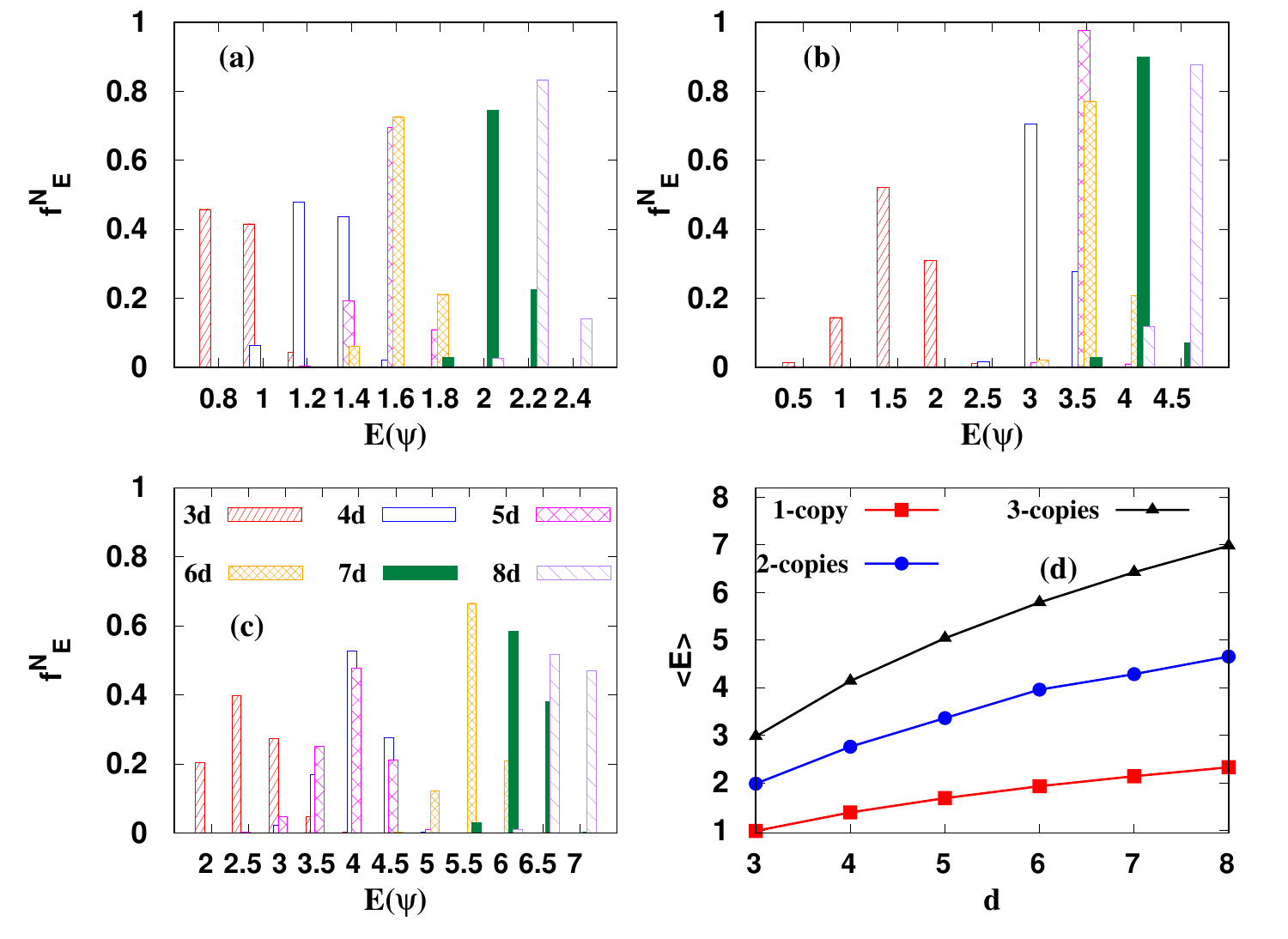}
		\caption{(Color Online.) Entanglement patterns of random pure bipartite states in $d \otimes d$. The normalized frequency distribution of entanglement (quantified by entanglement entropy), \(f_E^N\) (ordinate) is shown against \(E(\psi)\) (abscissa) for (a) $1$-copy, (b) $2$-copies and (c) $3$-copies. (d) Mean entanglement \(\langle E\rangle\) (ordinate) with respect to the dimension, d (abscissa) for $1$-copy (bottom line with squares), $2$-copies (middle line with circles) and $3$-copies (top line with triangles). We Haar uniformly simulate $10^6$ states in each dimension. All the axes are dimensionless.}
		\label{fig:ent_dist}
	\end{figure}
	
	Let us analyze  the patterns of the frequency  distribution in the entanglement content of pure  Haar uniformly generated bipartite states in different dimensions.
	It is known that the average multipartite entanglement of pure states increases with the increase of 
	the number of parties \cite{Gross_PRL_2009, Winter_PRL_2009, Aditi_PRA_2019} and hence it will be interesting to see whether such a trade-off is also true with the increase of dimension \cite{Kendon_PRA_2002}. 
	
In $d \otimes d$,  entanglement content of a pure state $|\psi \rangle_{AB}$ can be  quantified by the von Neumann entropy of the local density matrices \cite{Bennett_PRA_1996}, i.e., 
$E(|\psi\rangle_{AB}) = S(\rho_A)$ where $\rho_A$ is the reduced density matrix of the subsystem $A$ and  $S(\sigma) = - \text{Tr}(\sigma \log_2 \sigma)$. Since we will be dealing with two parties in this work, for notational simplicity, we will denote $E(|\psi \rangle_{AB})$ as  \(E(\psi)\) from now on. 

For a single copy,   the mean entanglement of an $M \otimes K$ bipartite system is predicted to be \cite{Kendon_PRA_2002} 
\begin{equation}
    \langle S_M \rangle \approx \log_2 M - \frac{M}{2K},
    \label{eq:approxvN}
\end{equation}
where \(S_M\) denotes the von Neumann entropy of \(\rho\) of dimension, \(M\). 
In our case, using $M = K = d$, we can easily estimate the average entanglement for single-copy of arbitrary two-qudit states
as demonstrated in Table. \ref{tab:entanglement_dist}. The normalized frequency distribution of entanglement,  \(f_E^N\), defined as \begin{equation}
    \frac{\mbox{Number of random states with a fixed entanglement, } E }{\mbox{Total number of simulated states with a given}, E},
    \label{eq:freqEdxd}
\end{equation}
is depicted  in Fig. \ref{fig:ent_dist}. Numerical simulations show that with the increase of  the dimension, the state becomes more and more entangled on average, 
which is in good agreement with the analytical estimates. Further, we analyse the entanglement when more than one copy of the state is taken into account. 
As expected, the average entanglement increases with the number of copies considered which can also be approximated via Eq. (\ref{eq:approxvN}).

Let us now investigate  the second moment of the distribution about mean, i.e., the standard deviation.  
Interestingly, we observe that for low number of copies, the states are concentrated around the mean entanglement, while as we consider $k \geq 2$, the distribution becomes broader  
which leads to the increase in standard deviation with the number of copies as elucidated in Table. \ref{tab:entanglement_dist}. 
We will show that the above analysis on entanglement patterns can help to examine 
the transformation properties between pair of states 
as discussed in the succeeding sections.

\begin{table}[]
			\resizebox{0.5\textwidth}{!}{\begin{minipage}{0.7\textwidth}
					\caption{Mean and standard deviation of the entanglement content of pure bipartite states with single and multiple copies in $d \otimes d$.} 
					\label{tab:entanglement_dist}
					\centering
					\begin{tabular}{|r|rrr|rr|rr|}
\hline
\multicolumn{1}{|c|}{{ d}} & \multicolumn{3}{c|}{{ 1-copy}}                                                                                                                                             & \multicolumn{2}{c|}{{ 2-copies}}                                                                       & \multicolumn{2}{c|}{{ 3-copies}}                                                                       \\ \hline
\multicolumn{1}{|c|}{{ }}  & \multicolumn{1}{c|}{{ }}                      & \multicolumn{1}{c|}{{ }}                    & \multicolumn{1}{c|}{{ }}             & \multicolumn{1}{c|}{{ }}                    & \multicolumn{1}{c|}{{ }}             & \multicolumn{1}{c|}{{ }}                    & \multicolumn{1}{c|}{{ }}             \\ \hline
\multicolumn{1}{|c|}{{ }}  & \multicolumn{1}{c|}{{ $\langle S_d \rangle$}} & \multicolumn{1}{c|}{{ $\langle E \rangle$}} & \multicolumn{1}{c|}{{ $E_{\sigma}$}} & \multicolumn{1}{c|}{{ $\langle E \rangle$}} & \multicolumn{1}{c|}{{ $E_{\sigma}$}} & \multicolumn{1}{c|}{{ $\langle E \rangle$}} & \multicolumn{1}{c|}{{ $E_{\sigma}$}} \\ \hline
\multicolumn{1}{|l|}{{ }}  & \multicolumn{1}{l|}{{ }}                      & \multicolumn{1}{l|}{{ }}                    & \multicolumn{1}{l|}{{ }}             & \multicolumn{1}{l|}{{ }}                    & \multicolumn{1}{l|}{{ }}             & \multicolumn{1}{l|}{{ }}                    & \multicolumn{1}{l|}{{ }}             \\ \hline
{ 3}                       & \multicolumn{1}{r|}{{ 1.09}}                  & \multicolumn{1}{r|}{{ 0.9930}}              & { 0.1309}                            & \multicolumn{1}{r|}{{ 1.9850}}              & { 0.3321}                            & \multicolumn{1}{r|}{{ 2.9770}}              & { 0.4932}                            \\ \hline
{ 4}                       & \multicolumn{1}{r|}{{ 1.5}}                   & \multicolumn{1}{r|}{{ 1.3816}}              & { 0.1149}                            & \multicolumn{1}{r|}{{ 2.7618}}              & { 0.1811}                            & \multicolumn{1}{r|}{{ 4.1420}}              & { 0.3412}                            \\ \hline
{ 5}                       & \multicolumn{1}{r|}{{ 1.82}}                  & \multicolumn{1}{r|}{{ 1.6821}}              & { 0.0975}                            & \multicolumn{1}{r|}{{ 3.3625}}              & { 0.1109}                            & \multicolumn{1}{r|}{{ 5.0429}}              & { 0.3946}                            \\ \hline
{ 6}                       & \multicolumn{1}{r|}{{ 2.08}}                  & \multicolumn{1}{r|}{{ 1.9319}}              & { 0.0845}                            & \multicolumn{1}{r|}{{ 3.8607}}              & { 0.1697}                            & \multicolumn{1}{r|}{{ 5.7927}}              & { 0.2537}                            \\ \hline
{ 7}                       & \multicolumn{1}{r|}{{ 2.3}}                   & \multicolumn{1}{r|}{{ 2.1437}}              & { 0.0740}                            & \multicolumn{1}{r|}{{ 4.2863}}              & { 0.1483}                            & \multicolumn{1}{r|}{{ 6.4289}}              & { 0.2243}                            \\ \hline
{ 8}                       & \multicolumn{1}{r|}{{ 2.5}}                   & \multicolumn{1}{r|}{{ 2.3296}}              & { 0.0657}                            & \multicolumn{1}{r|}{{ 4.6569}}              & { 0.1325}                            & \multicolumn{1}{r|}{{ 6.9834}}              & { 0.1981}                            \\ \hline
\end{tabular}
			\end{minipage}}
		\end{table}

\section{Typical property of bipartite entanglement transformation}
\label{sec:1}


We search for a universal pattern in LOCC covertability among typical pure states. 
In particular, 
we generate $10^6$ pairs of bipartite states, $|\psi\rangle_{AB}$ and $|\phi \rangle_{AB}$ with $E(\psi) \geq E(\phi)$, and calculate the distribution as well as percentage of comparable and incomparable pairs
in $d \otimes d$ (\(d = 3,4,5\)) (which will be denoted as $3$d, $4$d, $5$d) for $k = 1,2,3$ number of copies   i.e., for the transformation of $|\psi \rangle_{AB}^{\otimes k} \to |\phi \rangle_{AB}^{\otimes k}$ under LOCC. 

\begin{figure}[h]
		\centering
		\includegraphics[width=\linewidth]{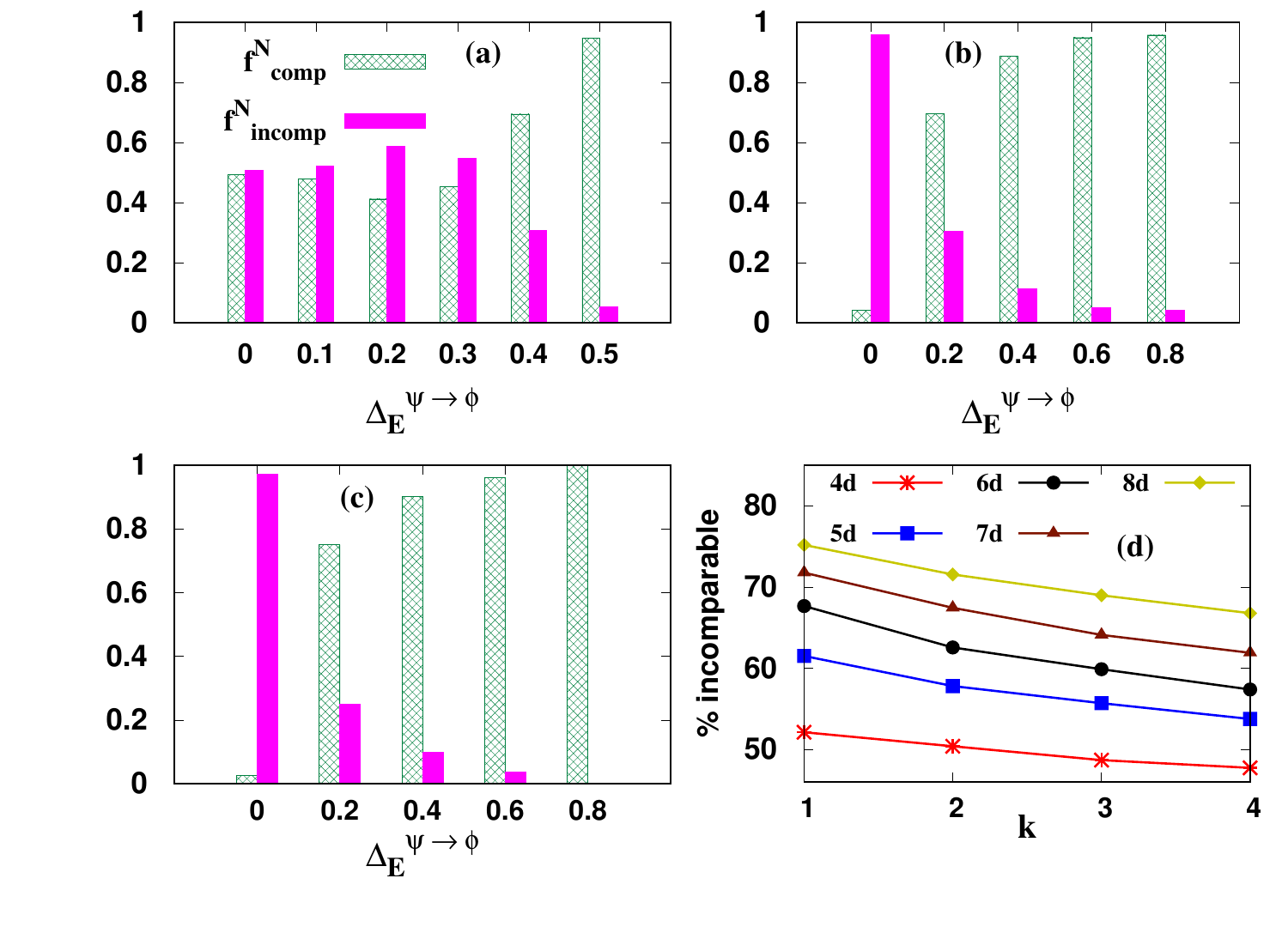}
		\caption{(Color Online.) Incomparable vs. comparable states.  Normalized frequency distributions of incomparable \(f^N_{incomp}\) (solids) and comparable states, \(f^N_{comp}\) (stripes) (vertical axis) vs. \(\Delta_E^{\psi \to \phi}\) (horizontal axis) when \(10^6\) states are generated Haar uniformly with \(E(\psi) \geq E(\phi)\) in (a) $3 \otimes 3$, (b) $4 \otimes 4$, and  (c) $5 \otimes 5$. 
		(d) The percentage of incomparable bipartite states (ordinate) with respect to the number of copies $k$ (abscissa) in $4$d (red stars), $5$d (blue squares), $6$d (black circles), $7$d (brown triangles) and $8$d (yellow diamonds). 	(In short, \(d\otimes d\) is referred to as \(d\)d  systems.)  All the axes are dimensionless. }
		\label{fig:incomp-comp}
	\end{figure}

First of all,  we observe that irrespective of the number of copies, the fraction of comparable pairs remains unaltered for two-qutrit states, i.e.,  once a pair of states is incomparable in $3 \otimes 3$, it remains so no matter how many copies of each state we consider as shown  in the literature \cite{Ujjwal_PRA_2002}. However, in higher dimensions, the number of incomparable pairs becomes more comparable with increasing number of copies although the percentage of multi-copy LOCC transformable pairs  decreases with the increase in the dimension for a fixed number of copies, as shown in Fig. \ref{fig:incomp-comp} (d).

For random source-target pairs, let us try to find whether there exists a connection between the incomparability (comparability) of the pairs  and  the difference in their entanglement content. The difference between the entanglement of the initial and the target states can be represented as 
\begin{equation}
    \Delta_E^{\psi \to \phi} = E(\psi) - E(\phi),
    \label{eq:Delta}
\end{equation}
for a given entanglement measure, \(E\),
when a single copy of \(|\psi\rangle_{AB}\) and \(|\phi\rangle_{AB}\) is available. When \(k\) copies of them are given,  the above definition can be modified as 
\begin{equation}
    \Delta_E^{k (\psi \to \phi)} = E(\psi^{\otimes k}) - E(\phi^{\otimes k}).
    \label{eq:Deltak}
\end{equation}
 To establish such connection between entanglement transformation and entanglement difference, we first numerically compute the normalized frequency distribution of incomparable (comparable) states for a given \(\Delta_E^{\psi \to \phi}\) in  \(d \otimes d\), which is defined as
\begin{equation}
    f^N_{incomp} =  \frac{\mbox{Number of incomparable states with a fixed}\, \Delta_E^{\psi \to \phi}}{\mbox{Total number of states simulated with a given}\, \Delta_E^{\psi \to \phi} }.
\end{equation}
Similarly, one can compute \(f^N_{comp}\) for comparable states found from the Haar uniformly generated states. Again for \(k\) copies, the frequency distribution can be redefined accordingly by using \(\Delta_E^{k (\psi \to \phi)}\).   As shown in Fig. \ref{fig:incomp-comp} (a)-(c), we observe that  when the difference in entanglement between the two states is very low, they tend to be incomparable which implies that the transformation between states with the same  entanglement content is rarely possible. However, as the two states differ more and more in their entanglement content, the number of comparable states increases. This feature becomes more prominent in higher dimensions. 
From the behavior of \(f^N_{incomp}\) in figure, we can safely claim that when \(\Delta_E^{k(\psi \to \phi)} >>0\), i.e., when $E(\psi^{\otimes k}) \gg E(\phi^{\otimes k})$, the percentage of comparable states increases drastically. 
Let us prove the above observation for qutrit pairs  
in the following theorem. 

\textbf{Theorem 1}. \textit{ 
 An increase in the difference in entanglement content between a pair of two-qutrit states indicates an increase in their comparability under LOCC in a single-copy level and vice-versa}.\\
  \textit{Proof}. Let us consider a pair of $3 \otimes 3$ states, $|\psi \rangle_{AB}$ and $|\phi \rangle_{AB}$, such that $E(\psi) \geq E(\phi)$, represented by ordered Schmidt coefficients as $\boldsymbol{\psi} \equiv \{\sqrt{a_1}, \sqrt{a_2}, \sqrt{a_3}\}$ and $\boldsymbol{\phi} = \{\sqrt{b_1}, \sqrt{b_2}, \sqrt{b_3}\}$ with $a_i \geq a_{i+1}$ and similarly for $b_i$
  and hence we consider the transformation $|\psi\rangle_{AB} \to_{LOCC} |\phi\rangle_{AB}$.  We quantify the entanglement of the states by their squared concurrence which are given by $C^2_{\psi} = 4 a_1 a_2 a_3$ and $C^2_{\phi} = 4 b_1 b_2 b_3$ \cite{Horodecki_RMP_2009}. We can drop the factor of $4$, since it does not affect the end results. Let $a_1 - b_1 = \alpha_1$ and $(a_1 + a_2) - (b_1 + b_2) = \alpha_2$. By Nielsen's theorem \cite{Nielsen_PRL_1999}, the two aforementioned states, $|\psi\rangle_{AB}$ and $|\phi\rangle_{AB}$, are incomparable if either $\alpha_1 > 0$ or $\alpha_2 >0$ or both are positive definite. The entanglement difference in this case can be represented as 
\begin{eqnarray}
\nonumber && \Delta_E^{\psi \to \phi}  = b_1 b_2 b_3 - a_1 a_2  a_3 \\
\nonumber && = b_1 b_2 b_3 - (\alpha_1 + b_1)(\alpha_2 - \alpha_1 + b_2)(b_3 - \alpha_2). \\
~~
\label{eq:ent_diff}
\end{eqnarray}
By assumption, $ \Delta_E^{\psi \to \phi} \geq 0$. 
Solving for $\frac{\partial \Delta_E^{\psi \to \phi}}{\partial{\alpha_1}} = 0$ and $\frac{\partial \Delta_E^{\psi \to \phi}}{\partial{\alpha_2}} = 0$, we obtain the following   values of $\alpha_1$ and $\alpha_2$, at which $\Delta_E^{\psi \to \phi}$ has an extremum, given by
\begin{equation}
    \alpha_1 = \frac{1}{3}(1 - 3b_1) ~~ \text{and} ~~ \alpha_2 = \frac{1}{3}(3b_3 - 1).
    \label{eq:alpha12max}
\end{equation}
We note that $\alpha_1 > 0$ implies that $b_1 < 1/3$, which is a contradiction since it is the highest Schmidt coefficient 
of $|\phi\rangle_{AB}$. Similarly, for $\alpha_2$ to be positive definite, we  need $b_3 > 1/3$, which again gives a contradiction since \(b_3\) is the lowest Schmidt coefficient. Thus both $\alpha_1, \alpha_2 \leq 0$, which implies that the states are comparable. Finally, we find that for such choices of $\alpha_1$ and $\alpha_2$, we have
\begin{eqnarray}
&&\frac{\partial^2 \Delta_E^{\psi \to \phi}}{\partial \alpha_1^2} . \frac{\partial^2 \Delta_E^{\psi \to \phi}}{\partial \alpha_2^2} - \Big(\frac{\partial^2 \Delta_E^{\psi \to \phi}}{\partial \alpha_1 \partial \alpha_2}\Big)^2 = \frac{1}{3} > 0  \label{eq:extremaE} \\
&& \text{and} ~~~\frac{\partial^2 \Delta_E^{\psi \to \phi}}{\partial \alpha_1^2} = \frac{\partial^2 \Delta_E^{\psi \to \phi}}{\partial \alpha_2^2} = -\frac{2}{3} < 0. \label{eq:maximaE}
\end{eqnarray}
which indicate a maximum in the entanglement difference. This is the only possible maximum in the entanglement difference, because the values of $\alpha_1$ and $\alpha_2$ in Eq. \eqref{eq:alpha12max} are the only possible ones which correspond to an extremum in Eq. \eqref{eq:ent_diff}. The other solutions for $\alpha_1$ and $\alpha_2$ corresponding to Eq. \eqref{eq:alpha12max} represent inflexion points. Thus $\Delta_E^{\psi \to \phi}$ can have only one maximum and Eqs. \eqref{eq:alpha12max}, \eqref{eq:extremaE} and  \eqref{eq:maximaE} demonstrate that if the entanglement difference between the two states is maximised,  they are comparable. This completes the first part of the proof.\\
Following a similar approach and writing  Eq. (\ref{eq:ent_diff}) in terms of \(\alpha_1\), \(\alpha_2\), \(a_1\), \(a_2\) and \(a_3\), we can show that $\Delta_E^{\psi \to \phi}$ attains a minimum for 
\begin{equation}
    \alpha_1 = \frac{1}{3}(3a_1 - 1) ~~ \text{and} ~~ \alpha_2 = \frac{1}{3}(1 - 3a_3).
    \label{eq:alpha12min}
\end{equation}
These values of $\alpha_1$ and $\alpha_2$ can only be positive definite, since otherwise we would have $a_1 < 1/3$ and $a_3 > 1/3$ which are contradictions since they are the maximum and minimum Schmidt coefficients characterising $|\psi \rangle_{AB}$. Thus, when the entanglement difference attains a minimum value (which again is unique, since these are the only values which represent an extremum of the entanglement difference), the states become incomparable. Since \(\Delta_E^{\psi \to \phi}\) is a smooth function, we can conclude that as the difference in entanglement between a pair of states increases (decreases), the states are more likely to be comparable (incomparable) i.e. their conversion probability increases (decreases). Hence the proof.
\hfill $\blacksquare$
	
\section{Connecting entanglement of catalyst and incomparable states}
\label{sec:2}

Having known the statistics of incomparable states in different dimensions, it is now important to study how other entangled states can facilitate their transformation and to find out whether there exists any relation between entanglement content of the catalysts and the difference in the entanglement of the pair of incomparable states. Towards that aim, 
for a given pair of incomparable states, \(|\psi \rangle_{AB}\) and $|\phi \rangle_{AB}$ which are also found from a random generation of \(5 \times 10^4\) states,  we Haar uniformly simulate \(10^5\) pure states and check whether they can act as catalysts or not. 



To make the analysis systematic, we introduce two quantifiers, mean and minimum entanglement of catalysts which will help us to establish the connection between entanglements of incomparable states and catalysts.  Let us consider a catalyst state $|\chi\rangle_{AB}$ in \(d\otimes d\), specified by its ordered Schmidt coefficients $\boldsymbol{\chi} = \{c_1,c_2, \cdots c_d\}$, in decreasing order which allows the transformation $|\psi \rangle_{AB} \to |\phi \rangle_{AB}$ to be executed via LOCC. The entanglement of the catalyst
  in terms of von Neumann entropy, \(E(\chi) = \sum_i c_i \log_2 c_i\)  is used in all the figures discussed in this paper.
The mean and minimum entanglement of catalysts are necessary to characterize the entanglement transformation since  for a given incomparable pair, there exist several catalyst states. The mean entanglement gives an idea about how much entanglement, on average, is required to facilitate a transformation otherwise impossible by LOCC
while the minimum entanglement of catalyst indicates the least additional resource required to mediate the transition.

Let the number of catalyst states found for a given incomparable pair be $\mathcal{N}$. For a fixed value of \(\Delta_E^{\psi \to \phi}\) of a given incomparable state with a single copy, we define the mean and minimum entanglement of catalysts as
\begin{eqnarray}
&& \langle E(\chi)\rangle = \frac{\sum_{i=1}^\mathcal{N} E(\chi_i)}{\mathcal{N}}, \, \, \mbox{and} \label{eq:cat_e_mean}  \\
&& E_{\min}(\chi) = \min_{i} E(\chi_i), \label{eq:cat_e_min}
\end{eqnarray}
where the minimization is over the entanglement content of the set of catalyst states. Numerically, we estimate the aforementioned equations as follows. Suppose we have a single pair of incomparable states, $|\psi\rangle_{AB}$ and $|\phi\rangle_{AB}$, defined by the ordered Schmidt coefficients $\boldsymbol{\alpha}= \{a_1, a_2,..., a_d\}$ and   $\boldsymbol{\beta} = \{b_1, b_2,..., b_d\}$ such that  $\boldsymbol{\alpha}$ is not majorised by $\boldsymbol{\beta}$ (i.e., $\boldsymbol{\alpha} \nprec \boldsymbol{\beta}$). For this particular pair, we Haar uniformly generate $10^5$ random states. Out of these random states, let there be $\mathcal{N}$ states which can be used as catalysts,  $\{|\chi_i\rangle\}$, each specified by an ordered set of Schmidt coefficients $\boldsymbol{\chi_i} = \{c_1,c_2,..., c_d\}$, which can facilitate LOCC transformability between the original incomparable states ($\Rightarrow \boldsymbol{\alpha} \otimes \boldsymbol{\chi} \prec \boldsymbol{\beta} \otimes \boldsymbol{\chi}$). Each such catalyst state $|\chi_i\rangle$ has a certain value of entanglement $E(\chi_i)$, and for the given incomparable pair, we find the average entanglement of all the $\mathcal{N}$ catalyst states which leads to Eq. \eqref{eq:cat_e_mean} and the minimum entanglement among the $\mathcal{N}$ number of $E(\chi_i)$s gives Eq. \eqref{eq:cat_e_min} for a given pair of incomparable states. The entire analysis is carried out for $5 \times 10^4$ incomparable pairs for a fixed value of \(\Delta_E^{(\psi \to \phi)}\).
 We can similarly define  \(\langle E(\chi)\rangle\) and \(E_{\min}(\chi)\) for a fixed value of \(\Delta_E^{k(\psi \to \phi)}\) with \(k\) copies being available.   Notice that these two quantifiers can extract two kinds of information about catalysts from the resource theoretic perspective -- average entanglement of the catalysts  quantifies their difficulty  in transformation while the minimum entanglement manifests how   less the assistance is enough to make the transformation possible.


Moreover, we wish to identify a set of incomparable states which are easier to catalyse, based on the entanglement content of incomparable pairs and hence we define the average number of catalyst states available within a given range of entanglement difference, \(\Delta_E^{\psi \to \phi}\), as 
\begin{eqnarray}
\langle n_\chi \rangle = \frac{\text{Number of catalysts found when } \Delta_E^{\psi \to \phi} \in (a,b)}{\text{Number of incomparable pairs having}~ \Delta_E^{\psi \to \phi} \in (a,b)}. \nonumber \\
\label{eq:cat_no}
\end{eqnarray}
Our main motivation for defining $\langle n_\chi \rangle$ above is to estimate the number of catalysts that exist for incomparable pairs within a particular entanglement difference. This gives us an idea about which entanglement range is more suitable for catalysts to exist. However, merely the number of catalysts within $\Delta_E^{\psi \to \phi} \in (a,b)$ gives a wrong interpretation of the  frequency distribution of catalysts. We have already seen from Theorem 1 that a lower entanglement difference leads to a larger number of incomparable pairs and vice versa. Hence a higher number of catalysts for a small value of $\Delta_E^{\psi \to \phi}$ has its origin in the existence of a large number of incomparable pairs i.e., since we only look for catalysts when states in a given pair are incomparable. Since the number of incomparable pairs is smaller for large entanglement differences, the number of catalysts found in that range of $\Delta_E^{\psi \to \phi} \in (a,b)$ is also smaller than that at $\Delta_E^{\psi \to \phi} \in (a,b) \to 0$. 
Hence defining $\langle n_\chi \rangle$ by considering only catalysts states when $\Delta_E^{\psi \to \phi} \in (a,b)$ is not proper. 
\\
To understand the distribution of catalysts within a given entanglement range, we must also consider the number of available incomparable pairs within that range of $\Delta_E^{\psi \to \phi} \in (a,b)$. Taking the number of incomparable pairs within $\Delta_E^{\psi \to \phi}$ into account, we define Eq. \eqref{eq:cat_no} such that it indicates the proper distribution of the number of catalysts available given a particular entanglement difference. A higher value of $\langle n_\chi \rangle$ implies that the number of catalyst states found with $\Delta_E^{\psi \to \phi} \in (a,b)$ is more than the number of incomparable pairs with a single copy in that range. This, in turn, indicates that such incomparable states are easily catalysed and the relation of $\langle n_\chi \rangle$  with \( \Delta_E^{\psi \to \phi} \in (a,b)\)  depicts the entanglement range where such states are prevalent. In a similar fashion, we can define \(\langle n_{\chi}^k\rangle\) with \(\Delta_E^{k(\psi \to \phi)}\).   Equipped with Eqs. \eqref{eq:Delta}, \eqref{eq:cat_e_mean}, \eqref{eq:cat_e_min} and \eqref{eq:cat_no}, let us proceed to the results in the single- and multiple-copy regimes.\\

 \begin{widetext}
	\begin{table}[]
			\resizebox{1.0\textwidth}{!}{\begin{minipage}{1.0\textwidth}
					\caption{Average and standard deviation of the mean and minimum  entanglement of catalysts for single- and multi-copy incomparable state transformation} 
					\label{tab:cata_ent}
					\centering
					\begin{tabular}{|c|crrr|crrr|}
\hline
{ }                        & \multicolumn{4}{c|}{{ \(4 \otimes 4\)}}                                                                                                                                                                                                                                           & \multicolumn{4}{c|}{{ \(5 \otimes 5\)}}                                                                                                                                                                                                                                           \\ \hline
{ }                        & \multicolumn{1}{c|}{{ }}                           & \multicolumn{1}{c|}{{ }}                 & \multicolumn{1}{c|}{{ }}                                 & \multicolumn{1}{c|}{{ }}                      & \multicolumn{1}{c|}{{ }}                           & \multicolumn{1}{c|}{{ }}                 & \multicolumn{1}{c|}{{ }}                                 & \multicolumn{1}{c|}{{ }}                      \\ \hline
{ k}                       & \multicolumn{1}{c|}{{ $\langle E (\chi) \rangle^{avg}$}} & \multicolumn{1}{c|}{{  $\langle E (\chi) \rangle^\sigma$}} & \multicolumn{1}{c|}{{ $E_{min}^{avg}$}} & \multicolumn{1}{c|}{{  $E_{min}^{\sigma}$}} & \multicolumn{1}{c|}{{ $\langle E (\chi) \rangle^{avg}$}} & \multicolumn{1}{c|}{{   $\langle E (\chi) \rangle^{\sigma}$}} & \multicolumn{1}{c|}{{ $E_{min}^{avg}$}} & \multicolumn{1}{c|}{{ $E_{min}^{\sigma}$}} \\ \hline
\multicolumn{1}{|l|}{{ }}  & \multicolumn{1}{l|}{{ }}                           & \multicolumn{1}{l|}{{ }}                 & \multicolumn{1}{l|}{{ }}                                 & \multicolumn{1}{l|}{{ }}                      & \multicolumn{1}{l|}{{ }}                           & \multicolumn{1}{l|}{{ }}                 & \multicolumn{1}{l|}{{ }}                                 & \multicolumn{1}{l|}{{ }}                      \\ \hline
\multicolumn{1}{|r|}{{ 1}} & \multicolumn{1}{r|}{{ 1.3613}}                     & \multicolumn{1}{r|}{{ 0.2689}}           & \multicolumn{1}{r|}{{ 0.9918}}                           & { 0.3536}                                     & \multicolumn{1}{r|}{{ 1.5943}}                     & \multicolumn{1}{r|}{{ 0.1793}}           & \multicolumn{1}{r|}{{ 1.1137}}                           & { 0.2687}                                     \\ \hline
\multicolumn{1}{|r|}{{ 2}} & \multicolumn{1}{r|}{{ 1.2498}}                     & \multicolumn{1}{r|}{{ 0.1846}}           & \multicolumn{1}{r|}{{ 0.6271}}                           & { 0.3021}                                     & \multicolumn{1}{r|}{{ 1.5682}}                     & \multicolumn{1}{r|}{{ 0.1343}}           & \multicolumn{1}{r|}{{ 0.9520}}                           & { 0.2187}                                     \\ \hline
\multicolumn{1}{|r|}{{ 3}} & \multicolumn{1}{r|}{{ 1.2334}}                     & \multicolumn{1}{r|}{{ 0.1827}}           & \multicolumn{1}{r|}{{ 0.8221}}                           & { 0.2842}                                     & \multicolumn{1}{r|}{{ 1.5626}}                     & \multicolumn{1}{r|}{{ 0.1357}}           & \multicolumn{1}{r|}{{ 0.6084}}                           & { 0.2177}                                     \\ \hline
\end{tabular}
			\end{minipage}}
		\end{table}
		\end{widetext}

\subsection{Catalysis for single-copy incomparable states}
\label{subsec:singlecopy}

\begin{figure}[h]
		\centering
		\includegraphics[width=\linewidth]{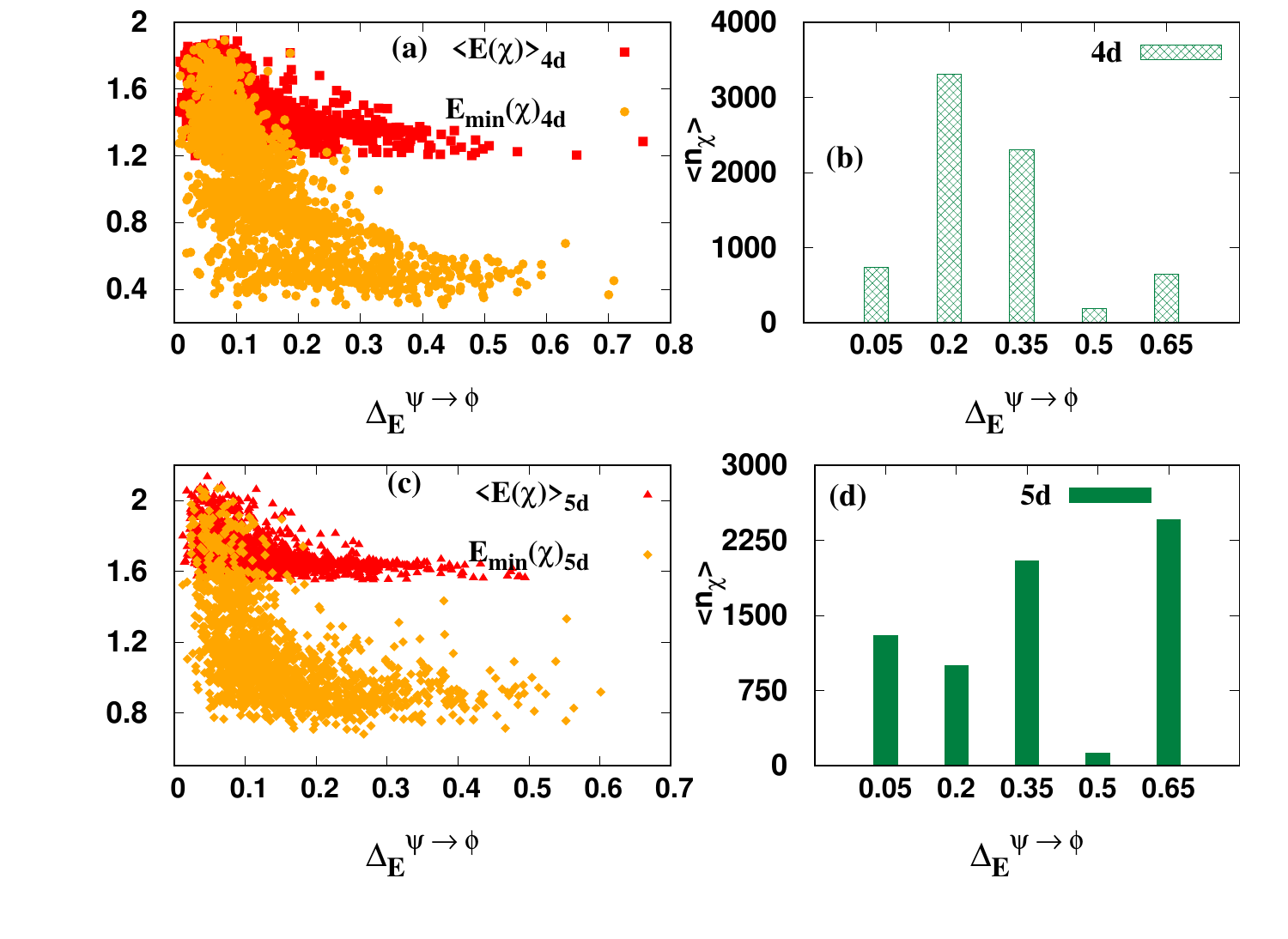}
		\caption{(Color Online.) 
		(a) Mean entanglement, $\langle E(\chi) \rangle$ (square)   and minimum entanglement,  $E_{min}(\chi)$  (circle) of catalyst states (ordinate) vs. \(\Delta_E^{\psi \to \phi}\) (abscissa) when random states  with \(E(\psi) \geq E(\phi)\) are used to check incomparability as well as for catalysts in \(4 \otimes 4\). (c) Triangles and diamonds represent  $\langle E(\chi) \rangle$  and   $E_{min}(\chi)$ respectively in \(5\otimes 5\).  (b) and (d) The average number of catalyst states, $\langle n_{\chi} \rangle$ (ordinate) against  \(\Delta_E^{\psi \to \phi}\)  (abscissa) for random states in $4$d (striped columns) (in (b)) and $5$d (solid columns) (in (d)). In all the plots,  states are incomparable when a single copy of the input-target pair is available.
		The data is computed by generating \(5 \times 10^4\) pairs to find incomparable states while for a given incomparable pair, \(10^5\) catalysts are simulated. 
		All the axes are dimensionless.
		}
		\label{fig:cata_1copy}
	\end{figure}
	
At the single-copy level, let us first investigate the trends of mean and minimum entanglement content of the catalyst required to facilitate the LOCC transformation between pair of incomparable states having  a fixed value of \( \Delta_E^{\psi \to \phi} \) (see  Fig. \ref{fig:cata_1copy}). 
For numerical simulation, we consider the catalyst dimension to be the same as that for the incomparable states. We observe that both the average and minimum entanglement of the catalyst states decrease with an increase in the entanglement difference of the incomparable states, \( \Delta_E^{\psi \to \phi}\).  It implies  that less entanglement is required as resource to expedite the transformation when the difference between the entanglement contents of incomparable pairs increases. 
On  one hand, when $E(\psi) \gg E(\phi)$, we are trying to convert a
highly entangled state to one with much less entanglement by LOCC. Since LOCC can reduce the entanglement content of the states on average, the decreasing trends of \(\langle E(\chi)\rangle\) or \(E_{\min}(\chi)\) is seen due to the fact that the transformations are readily implemented with only a little aid of the catalyst state. On the other hand, as we  argued in the previous section, the number of incomparable states decreases with the increase of \( \Delta_E^{\psi \to \phi}\), and even for the less number of incomparable states, a less amount of entanglement content is enough to make the transformation feasible. 
Furthermore, we notice that when the entanglement difference between an incomparable pair reaches its maximum value, the minimum entanglement which the catalyst should possess falls to almost half of the mean entanglement it can possess, thereby illustrating that
even states with meager entanglement content can augment transformation between pairs of incomparable states. 
Note also that these results do not depend on whether the individual states have very high or low entanglement content, but only the difference in their entanglement values matters. 
Therefore, our findings reveal that a catalyst required to transfer states with slightly different entanglement content may possess low entanglement as compared to the one which transforms two states having equivalent entanglement using LOCC.

 Let us now  focus on the average number of catalyst states that exist for a given pair of incomparable states with
 \( \Delta_E^{\psi \to \phi}\). As depicted in  Fig. \ref{fig:cata_1copy} (b) and (d) for states in $4 \otimes 4$ and $5 \otimes 5$ respectively, we notice that average number of catalysts   increases with the increase of \( \Delta_E^{\psi \to \phi}\). For example, in \(4 \otimes 4\), when  incomparable pairs are simulated with $ \Delta_E^{\psi \to \phi} \in (0.05, 0.1)$, $10^2$ catalysts  on average are found while  when \(\Delta_E^{\psi \to \phi} \in (0.7,0.75) \), the number of available catalysts shoots up drastically which is $\approx 3.14 \times 10^3$. A similar picture emerges for five dimensional bipartite states. It shows that  even though a very few incomparable states exist when $\Delta_E^{\psi \to \phi}$ is large,  several catalysts states are present which can help to make the pairs comparable. This result further points out how states having almost equal entanglement content are difficult to catalyse, even if their entanglement content is low.

		\subsection{Catalysis of multi-copy incomparable states}
\label{subsec:multiplecopy}
		
\begin{figure}[h]
		\centering
		\includegraphics[width=\linewidth]{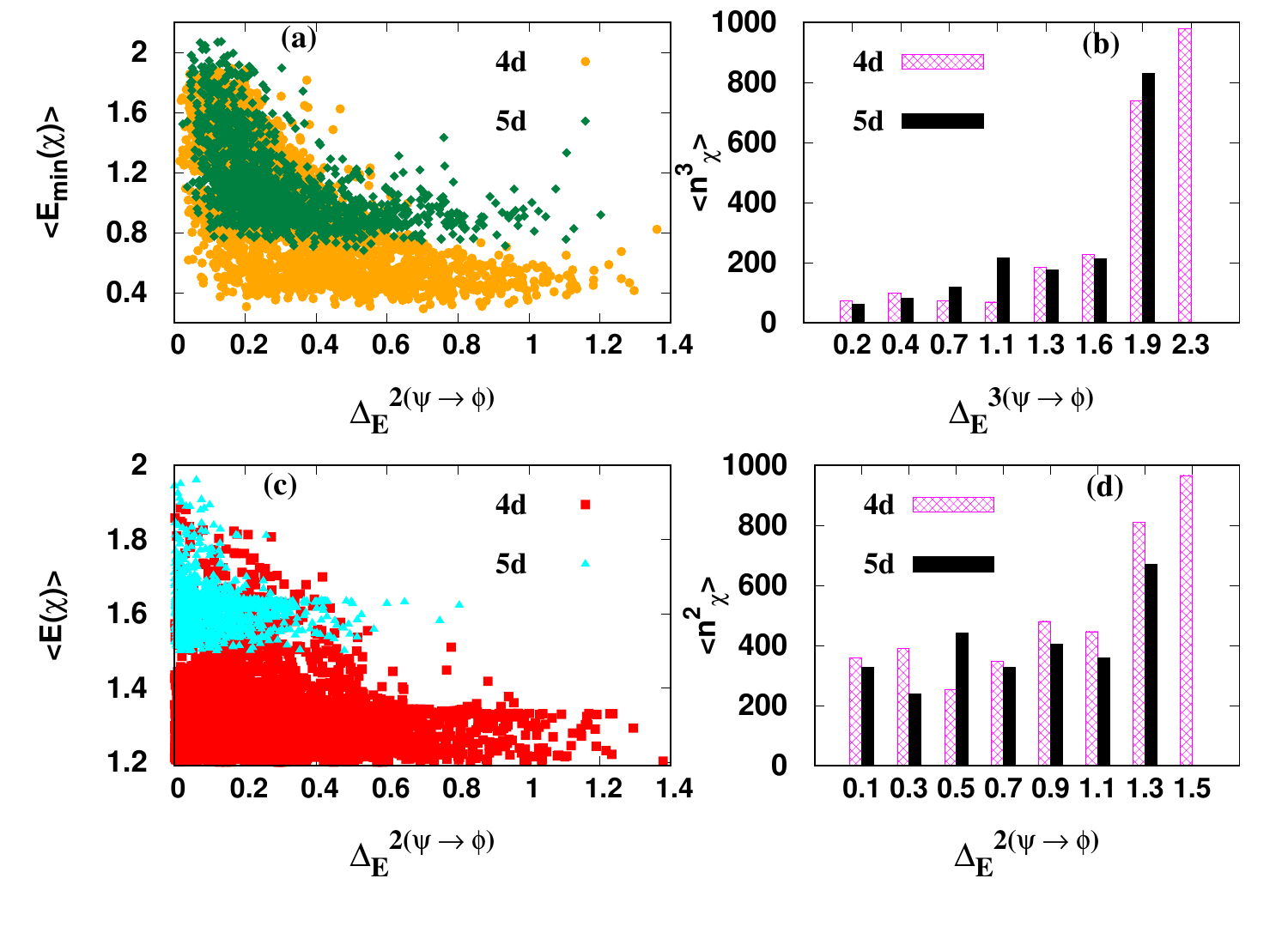}
		\caption{(Color Online.) 
		(a) \(E_{\min}(\chi)\)  of catalysts for $2$-copy incomparable states (ordinate) in $4d$ (circles) and $5d$ (diamonds)  with respect to $\Delta_E^{2(\psi \to \phi)}$ (abscissa). 
		(c) The same as in (a) except the ordinate is for $\langle E(\chi) \rangle$ (squares and triangles are used for states in \(4\)d and \(5\)d respectively). 
		(b) The average number of catalysts, $\langle n_{\chi}^3 \rangle$ for $3$-copy incomparable states  and (d) $\langle n_{\chi}^2 \rangle$ for $2$-copy incomparable states with $4$d (stripes) and $5$d (solid) (ordinate) against \(\Delta_E^{k(\psi \to \phi)}\). 
		 All the axes are dimensionless.
		}
		\label{fig:cata_multi}
	\end{figure}

		We now focus our attention on states which act as catalysts for $2$-copy and $3$-copy incomparable pairs, i.e., \(|\psi\rangle_{AB}^{\otimes k} \not\to_{LOCC} |\phi\rangle_{AB}^{\otimes k}\) with \(k=2, 3\) although 
		\(|\psi\rangle_{AB}^{\otimes k} |\chi\rangle \to_{LOCC} |\phi\rangle_{AB}^{\otimes k} |\chi\rangle\). 
		To investigate the patterns of such transformation, we first identify incomparable pairs in $2$-copy and $3$-copy level from randomly generated \(5 \times 10^4\) pairs having \(E(\psi) \geq E(\phi)\),  and then for a particular pair, among \(10^5\) randomly generated states, we find out the states which are suitable as catalysts.   The observations can be listed as follows. \\
		(1) Like a single-copy scenario, 
		we observe that both the mean and the minimum entanglement of catalyst decreases with the increase of 
		\(\Delta^{k(\phi \to \phi)}_E\) as illustrated in Fig. \ref{fig:cata_multi}.\\
		(2) The average number of available catalysts for a given incomparable pair also increases with an increase in \(\Delta^{k(\phi \to \phi)}_E\)  (\(k=2, 3\)).\\
		(3) Both \(\langle E(\chi)\rangle\) and \(E_{\min}(\chi)\) required to catalyse an incomparable pair increases with an increase in dimension. \\
		(4) The states which are incomparable at a higher number of copies need catalysts with a lower entanglement content to facilitate the transformation. This is because the additional copies of the states can also act as resources that can facilitate the transformation.
		
		To understand the dependence of the  entanglement in catalysts on the number of copies, we study the distribution of \(\langle E(\chi)\rangle\) and \(E_{\min}(\chi)\). Specifically, we calculate the average of the mean and the minimum entanglement of catalyst, i.e., 
		\begin{eqnarray}
		    \langle E(\chi)\rangle^{avg} & = & \frac{\sum_{k=1}^M \langle E(\chi)\rangle^k}{M}, \mbox{and} \nonumber \\
		     E_{\min}^{avg} & = & \frac{\sum_{k=1}^M E_{\min}^k(\chi)}{M},
		     \label{eq:cataavgsigma}
		\end{eqnarray}
		where \(M\) is the total number of incomparable states obtained from random state generation. In a similar fashion, we can compute the standard deviation of the distribution for the mean and the minimum entanglement of the catalyst states, denoted by 	\(\langle E(\chi)\rangle^{\sigma}\) and \(E_{\min}^{\sigma}\). The first moment and the second moment about the mean of the distribution for the entanglement content in the catalysts are calculated in	Table. \ref{tab:cata_ent}
		which reveals that the average as well as the standard deviation of the distribution for both the quantifiers decreases with the number of copies although both of them increase with the increase of dimension. The reduction in  standard deviation can be interpreted like the following:  in entanglement transformation via LOCC, when two copies of source and target states are provided, it can be thought of as an additional resource given to the transformation and hence a set containing pairs which are two- or more-copy incomparable are harder to transform via LOCC  compared to a set of pairs which are single-copy incomparable and two- or more copy comparable. It also provides a classification among incomparable states. In that sense, the reduction of standard deviation of \(\langle E(\chi)\rangle\) and \(E_{\min}\)  implies that these pairs of two- or more copy incomparable states require much more stringent bounds on entanglement to catalyze. 
		Furthermore, comparing Figs. \ref{fig:cata_multi} (b) and (d) with Fig. \ref{fig:cata_1copy}, we realize that  for incomparable pairs with a given entanglement difference, the average number of catalysts available decreases with an increase in the number of copies. It can again be explained by noticing that such catalysts exist within a smaller region of the state space (as indicated by the lower standard deviation), and hence are harder to encounter.

	\section{Hierarchy of catalyst}
\label{sec:cat_hie}

Up to now, we studied characteristics of the catalyst states with respect to the entanglement contents of the input and output states generated randomly for LOCC transformation, thereby providing a classification among incomparable states. 
We now move on to characterize catalyst states according to their transformation power, which facilitates transformations of different kinds of incomparable pairs, thereby establishing a hierarchy among the catalysts. To establish the rankings among catalysts, we consider the following three situations. 
\begin{enumerate}
    \item \textit{One step-assisted catalysts.} Let us consider a pair of states, $|\psi\rangle_{AB}$ and $|\phi\rangle_{AB}$, such that $|\psi \rangle_{AB} \nrightarrow |\phi \rangle_{AB}$, as well as $|\psi \rangle_{AB}^{\otimes 2} \nrightarrow |\phi \rangle_{AB}^{\otimes 2}$ via LOCC. We investigate the entanglement properties of catalysts which can make the transformation possible at the single-copy level. We denote such catalyst states as $|\chi \rangle_{2(1)}$ where in the subscript, \(2\) represents the fact that  it can catalyze states that are $2$-copy incomparable while \((1)\) shows that it can make  pairs of states comparable in a single-copy level. 
    In general, if a pair is \(n+1\)-copy incomparable and if a catalyst can make it comparable in the \(n\)-copy level, the catalyst state can be denoted as \(|\chi\rangle_{n+1(n)}\) which can be called one step-assisted catalyst.
    In our study, we also investigate catalysts which can aid transformations at the $2$-copy level for pairs which are $3$-copy incomparable. We then compare such catalysts, deemed as $|\chi \rangle_{3(2)}$, with  $|\chi \rangle_{2(1)}$ and compare their entanglement properties and distributions with respect to each other. Notice that the set of catalysts found here is different than the one considered in the preceding section where if a pair is $2$-copy incomparable, we find  catalysts which make the pair comparable at the two-copy level itself. In general, \(m\) step-assisted catalysts, \(|\chi\rangle_{m(n)}\) with \(m \geq n\), can be introduced, where a pair is incomparable with \(m\) copies while with the help of catalysts, a pair becomes comparable with \(n\) copies.  
    
     \item \textit{Strong catalysts.} We now identify catalysts which can make a single-copy transformation possible between pairs although the pairs are incomparable at a higher number of copies. Specifically, we  generate \(10^5\) two-qudit states Haar uniformly and find the set of pairs which are incomparable at a high number of copies ($n \geq 3)$.  For these pairs, we determine the catalysts which can make the pairs comparable even at a single-copy level. We call these catalyst states as \textit{strong catalysts} and denote them as \(|\chi \rangle_{n(1)}\) whose entanglement features will be investigated. 
    
    \item  \textit{Cost-efficient catalysts.} Let us now categorize catalyst states which not only facilitate  LOCC impossible transformations, but also provide a lower resource cost in doing so. To determine them, let us consider $|\psi\rangle_{AB}$ and $|\phi\rangle_{AB}$ which are incomparable for a single-copy level although $|\psi \rangle_{AB}^{\otimes 2} \rightarrow_{LOCC} |\phi \rangle_{AB}^{\otimes 2}$. This implies that an extra copy of the initial state acts as an additional resource which can make the transformation feasible. For these pairs of states, we find catalyst states, denoted by $|\chi \rangle_2$, with $E(\chi_2) \leq E(\psi)$ and \(E(\chi_2) \leq E(\phi)\) which can make the single-copy transformation possible, thereby illustrating the advantage of using catalysts from the perspective of resource  instead of two copies of the pairs. 

    
    \end{enumerate}
    
     Our aim is to characterize all such sets of catalyst states generated Haar uniformly  in $4\otimes 4$ and $5 \otimes 5$ for a given incomparable pair. It will be interesting to find whether such sets possess certain entanglement patterns.

\subsection{Strong catalysts vs. one step-assisted catalysts}
\label{subsec:cat_hie_1}

\begin{figure}[h]
		\centering
		\includegraphics[width=\linewidth]{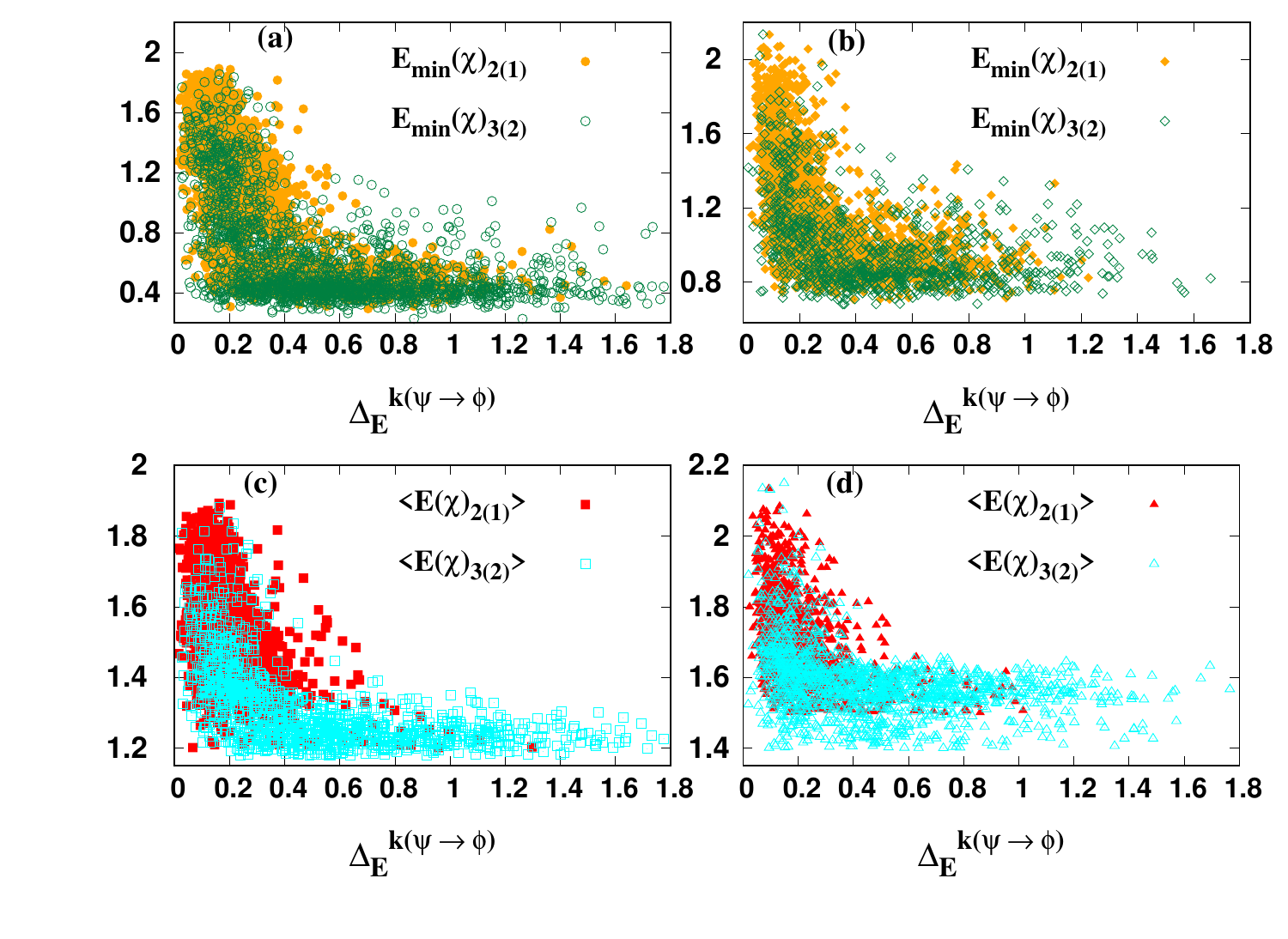}
		\caption{(Color Online.) Entanglement patterns of one step-assisted catalysts. (a) \( E_{\min} (\chi)_{2(1)}\)  (solid circles) and \(E_{\min} (\chi)_{3(2)}\)  (hollow circles) (ordinate) with respect to \(\Delta_E^{k(\psi \to \phi)}\) (abscissa) having \(k=1, 2\) respectively for randomly generated states in \(4 \otimes 4\). (c) The similar plot as in (a) for \(\langle E (\chi)_{2(1)}\rangle\) (solid squares) and \(\langle E (\chi)_{3(2)}\rangle\) (hollow squares). (b) and (d) are for random states  in  \(5 \otimes 5\). (b) is similar to (a) with  circles being replaced by  diamonds while (d) is similar to (c) except squaress are changed to triangles. All the axes are dimensionless.  
		}
		\label{fig:cata_hie}
	\end{figure}
	
To analyze the power of the catalysts, $|\chi\rangle_{n+1(n)}$, 
we first Haar uniformly generate \(5 \times 10^4\) states in $4 \otimes 4$ and $5 \otimes 5$, identify \(2\)- and \(3\)-copy incomparable pairs, and finally find states which can act as catalysis in a single- and two-copy levels respectively from another Haar uniform set of \(10^5\) states. We again use three quantifiers, 
\(\langle E(\chi)_{n+1(n)} \rangle\), \(E_{\min}(\chi)_{n+1(n)}\) and \(\langle n_{\chi_{n+1(n)}}\rangle\) to manifest their overall patterns with
\(\Delta_E^{n(\psi \to \phi)}\). Notice that in Sec. \ref{subsec:multiplecopy}, the behavior of catalysts are shown with respect to \(\Delta_E^{n+1(\psi \to \phi)}\).  

From Fig. \ref{fig:cata_hie}, we find that the mean entanglement of $|\chi \rangle_{2(1)}$ is higher that  of $|\chi \rangle_{3(2)}$ for   LOCC incomparable pairs in $4 \otimes 4$ and $5 \otimes 5$. This indicates that to catalyse incomparable pairs at the two-copy level, we require less amount of entanglement than to do so at the single-copy level, even though in both the cases the pair is incomparable at a higher number of copies. It also shows that the catalyst states are more resourceful than the additional copies of the initial states themselves. 
Again like multiple-copy transformation seen in the previous section, Table \ref{tab:cata_hie}  elucidates how
the standard deviation defined in Eq. (\ref{eq:cataavgsigma}) of the mean and minimum entanglement distribution decreases with the number of copies, thereby showing that a narrower entanglement range is available to choose the catalyst states from. 

\begin{figure}[h]
		\centering
		\includegraphics[width=\linewidth]{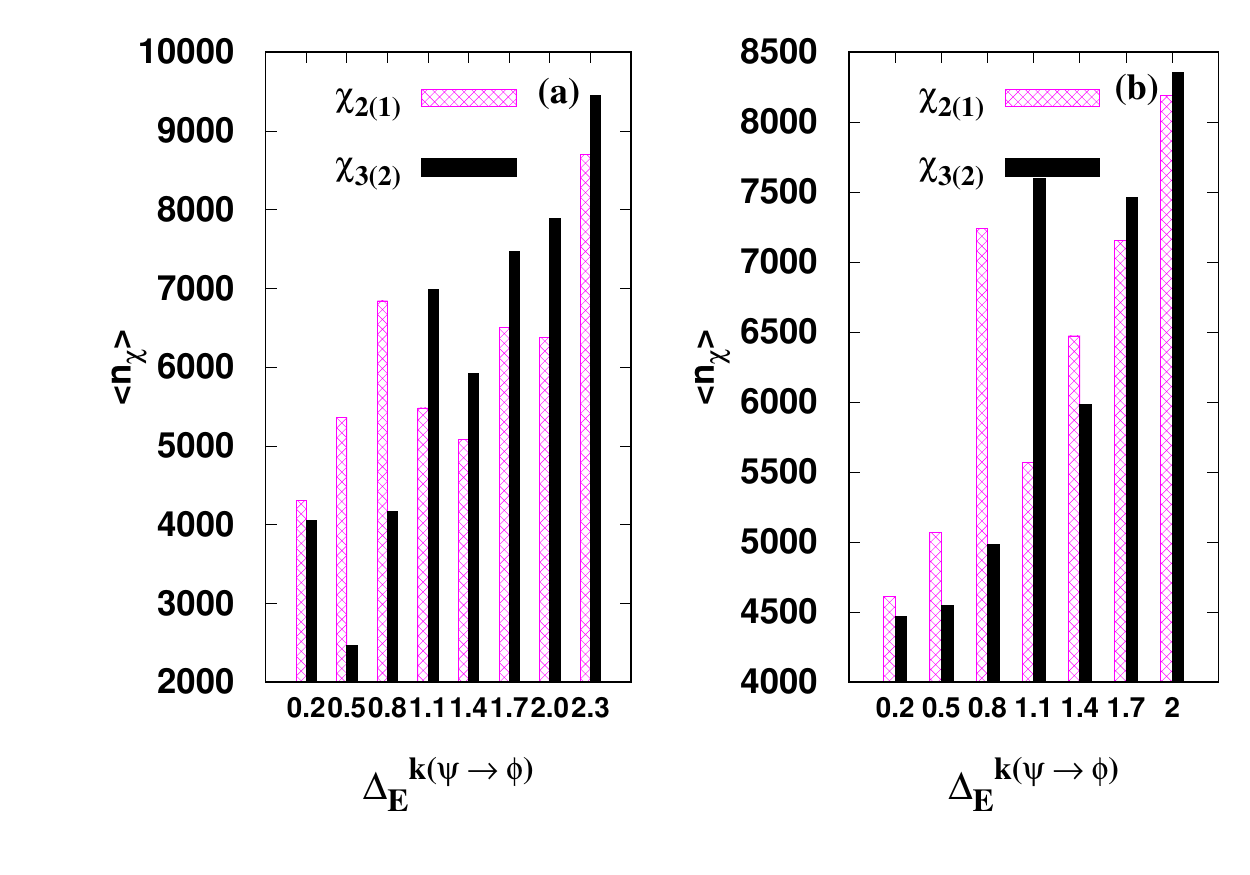}
		\caption{(Color Online.)  $\langle n_\chi \rangle$ (ordinate) against \(\Delta_E^{k(\psi \to \phi)}\) (abscissa) with \(k=1, 2\)  for catalysts $|\chi \rangle_{2(1)}$  (stripes) and $|\chi \rangle_{3(2)}$ (solid) respectively. (a) Distribution of $\langle n_\chi \rangle$ in $4 \otimes 4$ while (b) is for  $5 \otimes 5$. All the axes are dimensionless.  }
		\label{fig:cata_hie_no}
	\end{figure}

Comparing Figs. \ref{fig:cata_multi} and \ref{fig:cata_hie_no} for \(\langle n_{\chi} \rangle\), 
another feature of the catalyst hierarchy  emerges -- the average  number of catalysts, \(\langle n_{\chi_{3(2)}} \rangle\) which make $2$-copy LOCC transformation of $3$-copy incomparable states possible, increases when the entanglement difference of the incomparable pair, i.e., \(\Delta_E^{2(\psi \to \phi)}\) also increases, thereby showing monotonic behavior with \(\Delta_E^{2(\psi \to \phi)}\).  On the other hand, such a monotonic behavior is absent for \(\langle n_{\chi_{2(1)}} \rangle\) with \(\Delta_E^{\psi \to \phi}\) (see Fig. \ref{fig:cata_hie_no}). 
It is possibly due to the fact that  pairs which are \(3\)-copy incomparable are harder to catalyse. \\

\textbf{Remark.} In Fig. \ref{fig:cata_hie_no}, we plot \(\langle n_{\chi_{2(1)}} \rangle\) and \(\langle n_{\chi_{3(2)}} \rangle\) against the entanglement difference of the corresponding $k$ copy incomparable pair. It is observed that \(\langle n_{\chi_{2(1)}} \rangle\) is significantly more than \(\langle n_{\chi_{3(2)}} \rangle\) at low values of the entanglement difference. Specifically, we find that for $|\chi\rangle_{3(2)}$, the frequency distribution $\langle n_{\chi_{3(2)}} \rangle$ increases with the increase of difference in entanglement  much more prominently than that for $|\chi\rangle_{2(1)}$. By “monotonic”, we thus mean that the distribution of $|\chi\rangle_{3(2)}$ increases with the increase in  entanglement-difference which is not the case for $|\chi\rangle_{2(1)}$.
The number of $|\chi\rangle_{3(2)}$ is much less at low values of the entanglement difference and grows steadily with $\Delta_{E}^{k(\psi\to\phi)}$, although with minor fluctuations. Thus we can say that \(\langle n_{\chi_{3(2)}} \rangle\) is roughly monotonic with respect to the \(\Delta_E^{k(\psi\to\phi)}\), as compared to \(\langle n_{\chi_{2(1)}} \rangle\).
	\begin{figure}[h]
		\centering
		\includegraphics[width=\linewidth]{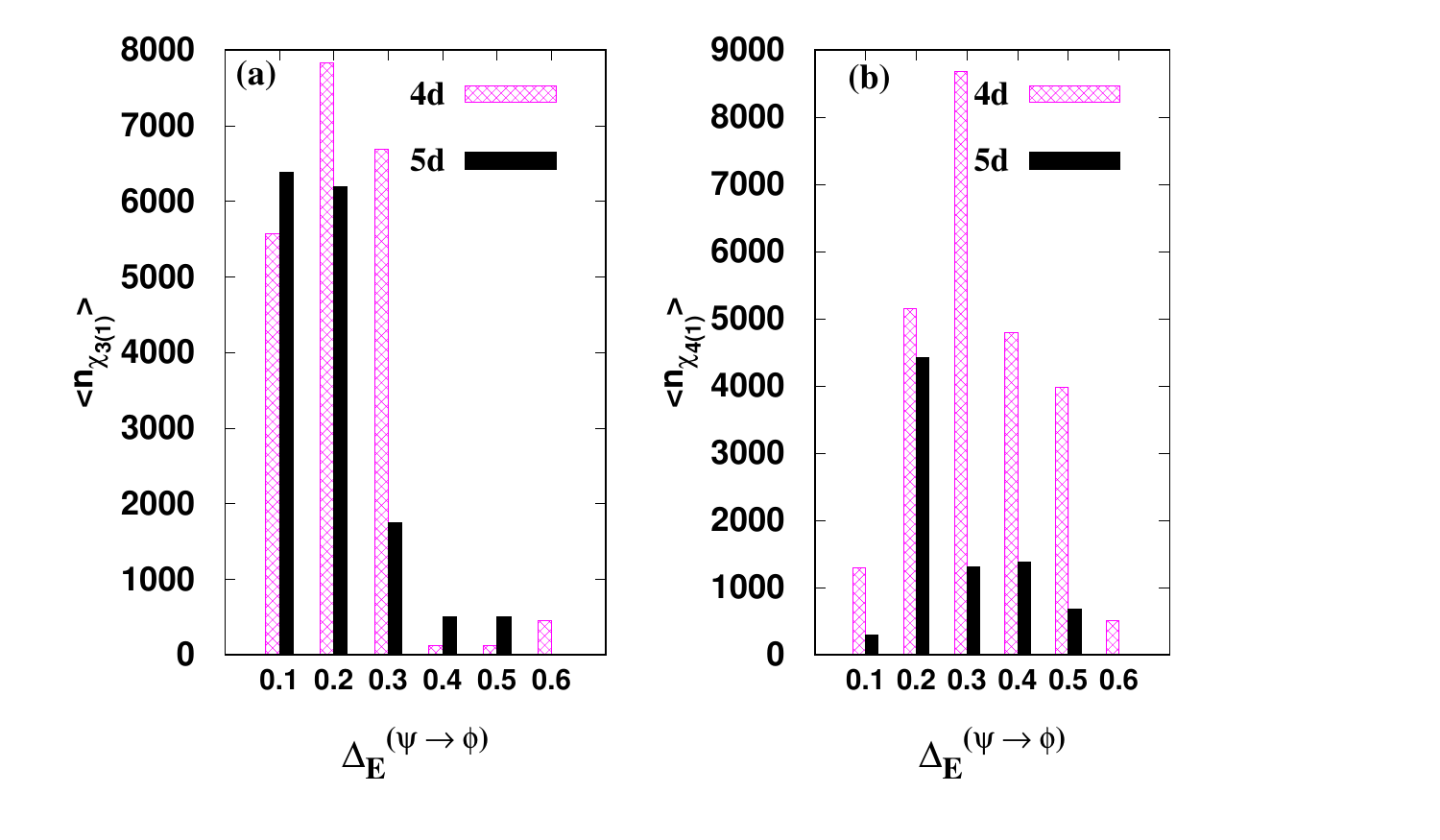}
		\caption{(Color Online.) Frequency distribution of average number of \textit{ strong catalysts}, \(\langle n_{\chi_{n(1)}} \rangle\)  (ordinate) with \(\Delta_E^{\psi \to \phi}\) (abscissa)  found  in $4\otimes4$ (stripes) and $5\otimes5$ (solid). (Left panel) (a) Strong catalysts \(\langle n_{\chi_{3(1)}} \rangle\) for $3$-copy incomparable states. (Right panel) (b) Strong catalysts \(\langle n_{\chi_{4(1)}} \rangle\) for $4 $-copy incomparable states. Unlike one step-assisted catalysts, they can be found in good numbers for low values of \(\Delta_E^{\psi \to \phi}\). All the axes are dimensionless. }
		\label{fig:cata_hie_no_for1}
	\end{figure}

In this respect, we can argue that $|\chi \rangle_{2(1)}$ are stronger catalysts than $|\chi \rangle_{3(2)}$  because the former can facilitate transformation between two highly entangled states while the latter requires the final state to be much less entangled so that the difference between entanglements in the source and the target states are high. For example, from Fig. \ref{fig:cata_hie_no}, we observe that for entanglement difference up to \(0.75\), \(\langle n_{\chi_{2(1)}} \rangle\) is greater than \(\langle n_{\chi_{3(2)}} \rangle\) which indicates that $|\chi\rangle_{2(1)}$ can catalyse incomparable states with small  difference in entanglement, i.e., they can catalyse transformations between two highly entangled states too. On the other hand, \(\langle n_{\chi_{3(2)}} \rangle\) is more significant at higher values of \(\Delta_E^{k(\psi\to\phi)}\) which means that $|\chi\rangle_{3(2)}$ are successful as catalysts for transformations between highly entangled and low entangled states. For this reason, we term $|\chi\rangle_{2(1)}$ as “stronger” catalysts since they may be able to facilitate transformations between two highly entangled incomparable states. Moreover, as the dimension of the bipartite states increases, the average number of catalysts goes down, as is evident by comparing  Fig. \ref{fig:cata_hie_no}(a) and (b).
	
		\begin{figure}[h]
		\centering
		\includegraphics[width=\linewidth]{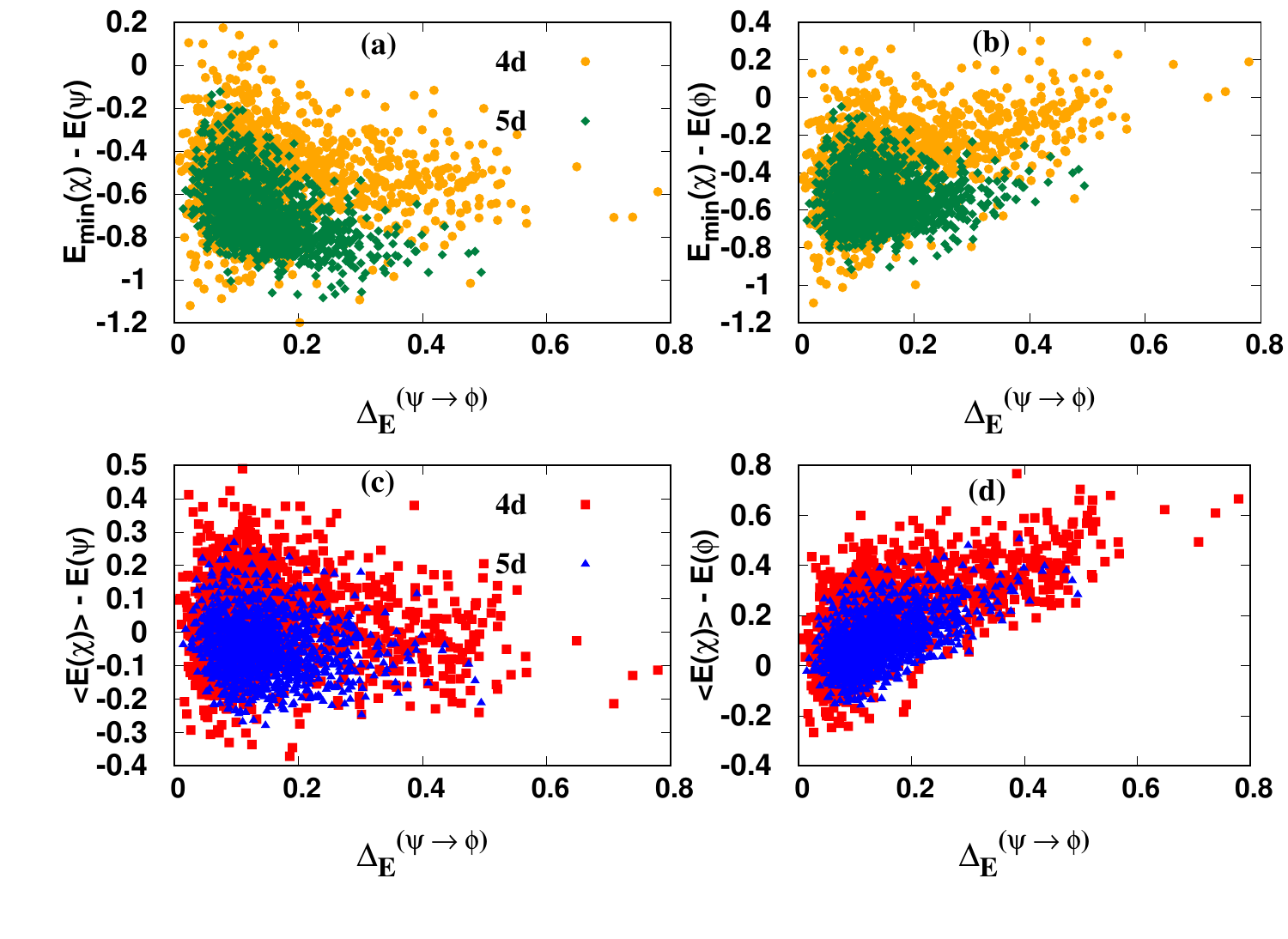}
		\caption{(Color Online.) Quantifying cost-efficient catalysts.  $ E_{\min}(\chi) - E(\psi)$  (in (a)) and $ E_{\min}(\chi) - E(\phi)$   (in (b)) (ordinate) against the single-copy entanglement difference,  $\Delta_E^{\psi \to \phi}$ (abscissa). Circles and diamonds represent the Haar uniformly generated states in \(4 \otimes 4\) and \(5\otimes 5\) respectively. (c) and (d) Similar plots as in (a) and (b) where the vertical axis is for $\langle E(\chi)  \rangle - E(\psi)$ and $\langle E(\chi) \rangle  - E(\phi)$ in (c) and (d) respectively. Squares and triangles correspond to \(4\)d and \(5\)d catalyst states. All the axes are dimensionless.} 
	
		\label{fig:cata_diff}
	\end{figure}


\textit{Strong catalysts.} Let us now explore the properties of catalyst states, \(|\chi\rangle_{n(1)}\) which can enable entanglement transformation  at the single-copy level even when multiple copies of the initial and final states remain incomparable. To investigate this, we randomly generate states which are incomparable under LOCC even when \(3\) and \(4\) copies are provided although by using catalysis, we can make the pair comparable with a single-copy. First of all, we notice that \(\langle E\rangle^{avg}\) and \(E_{\min}^{avg}\) increases with \(n\), thereby showing that such strong catalysts should possess a higher amount of entanglement compared to the one step-assisted catalysis (see Table \ref{tab:cata_hie} ). Secondly, as $n$ increases, the corresponding catalysts exist mainly when the entanglement difference between the states, \(\Delta_E^{\psi \to \phi}\) is low as shown in Fig. \ref{fig:cata_hie_no_for1}.
Furthermore, the average number of such \textit{strong catalysts} decreases with the increase of dimension. 

		\subsection{Cost-efficient catalysis}
\label{subsec:cat_hie_2}

To determine the set of cost-efficient catalysts, we look for catalysts for which $E(\chi) < E(\psi)$ as well as $ E(\chi) <E(\phi)$ hold so that  we can conclude that catalyst state is a less costly resource to furnish the desired transformation than the additional copy of the input states.
As depicted in Fig. \ref{fig:cata_diff}, we plot the difference between the mean (minimum) entanglement of catalysts  and \(E(\psi)\) as well as \(E(\phi)\) involved in the transformation. If the difference is negative,  we can argue that catalytic transformation is more beneficial than the transformation with two copies. For  $2$-copy comparable states both in $4 \otimes 4$ and $5 \otimes 5$, we find that a large fraction of the catalysts has lower entanglement than the initial and final states, and yet can help in converting a single copy of $|\psi\rangle_{AB}$ to $|\phi \rangle_{AB}$. For a pair of states admitting multiple catalysts, the mean entanglement required to aid the transition becomes negative as the entanglement difference between the states $E(\psi) - E(\phi)$ increases. The maximum amount of entanglement that can be saved as a resource is quantified by $E_{min}(\chi) - E_{min}(\psi)$ and it is shown that this quantity too becomes progressively negative with the increase in \(\Delta_E^{\psi \to \phi}\). Thus catalyst states not only help further an otherwise impossible LOCC transformation, but also act as a more efficient resource to do so, since they consume less entanglement than a second copy of the states. When compared with $E(\phi)$, we find that there still exist a large number of less entangled catalysts, although they decrease in number with growing \(\Delta_E^{\psi \to \phi}\). However, in our simulation, $|\psi\rangle_{AB}$ is taken to be  more entangled than $|\phi\rangle_{AB}$, and hence  catalysts having less entangled than $|\psi\rangle_{AB}$ are enough to highlight their efficiency as a resource. Furthermore, as we increase the dimension, the catalysts become less and less costly as compared to the initial states and thereby certify their beneficial role in the transformation protocol. Our data suggests that instead of using multiple copies of the same states, the search for a suitable catalyst can prove to be a more effective routine when one considers the entanglement that is ultimately consumed in the transformation protocol.

\section{Conclusion}
\label{sec:conclu}

\begin{widetext}
\begin{table}[]
			\resizebox{0.89\textwidth}{!}{\begin{minipage}{1.0\textwidth}
					\caption{Mean and standard deviation of the average and minimum  entanglement of \textit{one step-assisted}, \(|\chi\rangle_{2(1)}\), \(|\chi\rangle_{3(2)}\) and \textit{strong} catalysts, \(|\chi\rangle_{3(1)}\), $|\chi\rangle_{4(1)}$.} 
					\label{tab:cata_hie}
					\centering
					\begin{tabular}{|l|llll|llll|clll|lllll|}
\hline
\multicolumn{1}{|c|}{d} & \multicolumn{4}{c|}{$|\chi\rangle_{2(1)}$}                                                                                                                                            & \multicolumn{4}{c|}{$|\chi\rangle_{3(2)}$}                                                                                                                                            & \multicolumn{4}{c|}{$|\chi\rangle_{3(1)}$}                                                                                                                                                        & \multicolumn{5}{c|}{$|\chi\rangle_{4(1)}$}                                                                                                                                                                                      \\ \hline
\multicolumn{1}{|c|}{}  & \multicolumn{1}{c|}{}                                 & \multicolumn{1}{c|}{}                                    & \multicolumn{1}{c|}{}                & \multicolumn{1}{c|}{}       & \multicolumn{1}{c|}{}                                 & \multicolumn{1}{c|}{}                                    & \multicolumn{1}{c|}{}                & \multicolumn{1}{c|}{}       & \multicolumn{1}{c|}{}                                 & \multicolumn{1}{l|}{}                                    & \multicolumn{1}{l|}{}                &                                         & \multicolumn{1}{c|}{}                                 & \multicolumn{1}{c|}{}                                    & \multicolumn{1}{c|}{}                & \multicolumn{1}{c|}{}                   &                             \\ \hline
                        & \multicolumn{1}{l|}{$\langle E (\chi) \rangle^{avg}$} & \multicolumn{1}{l|}{$\langle E (\chi) \rangle^{\sigma}$} & \multicolumn{1}{l|}{$E_{min}^{avg}$} & $E_{min}^{\sigma}$          & \multicolumn{1}{l|}{$\langle E (\chi) \rangle^{avg}$} & \multicolumn{1}{l|}{$\langle E (\chi) \rangle^{\sigma}$} & \multicolumn{1}{l|}{$E_{min}^{avg}$} & $E_{min}^{\sigma}$          & \multicolumn{1}{c|}{$\langle E (\chi) \rangle^{avg}$} & \multicolumn{1}{c|}{$\langle E (\chi) \rangle^{\sigma}$} & \multicolumn{1}{c|}{$E_{min}^{avg}$} & \multicolumn{1}{c|}{$E_{min}^{\sigma}$} & \multicolumn{1}{c|}{$\langle E (\chi) \rangle^{avg}$} & \multicolumn{1}{c|}{$\langle E (\chi) \rangle^{\sigma}$} & \multicolumn{1}{c|}{$E_{min}^{avg}$} & \multicolumn{1}{c|}{$E_{min}^{\sigma}$} &                             \\ \hline
                        & \multicolumn{1}{l|}{}                                 & \multicolumn{1}{l|}{}                                    & \multicolumn{1}{l|}{}                &                             & \multicolumn{1}{l|}{}                                 & \multicolumn{1}{l|}{}                                    & \multicolumn{1}{l|}{}                &                             & \multicolumn{1}{l|}{}                                 & \multicolumn{1}{l|}{}                                    & \multicolumn{1}{l|}{}                &                                         & \multicolumn{1}{l|}{}                                 & \multicolumn{1}{l|}{}                                    & \multicolumn{1}{l|}{}                & \multicolumn{1}{l|}{}                   &                             \\ \hline
\multicolumn{1}{|r|}{4} & \multicolumn{1}{r|}{1.3423}                           & \multicolumn{1}{r|}{0.3329}                              & \multicolumn{1}{r|}{1.0674}          & \multicolumn{1}{r|}{0.4118} & \multicolumn{1}{r|}{1.2114}                           & \multicolumn{1}{r|}{0.2124}                              & \multicolumn{1}{r|}{0.6371}          & \multicolumn{1}{r|}{0.3365} & \multicolumn{1}{r|}{1.4396}                           & \multicolumn{1}{r|}{0.3117}                              & \multicolumn{1}{r|}{1.3001}          & \multicolumn{1}{r|}{0.2301}             & \multicolumn{1}{l|}{1.5257}                           & \multicolumn{1}{l|}{0.2896}                              & \multicolumn{1}{l|}{1.4787}          & \multicolumn{1}{l|}{0.1759}             & \multicolumn{1}{r|}{0.1759} \\ \hline
\multicolumn{1}{|r|}{5} & \multicolumn{1}{r|}{1.5523}                           & \multicolumn{1}{r|}{0.2187}                              & \multicolumn{1}{r|}{1.1800}          & \multicolumn{1}{r|}{0.3056} & \multicolumn{1}{r|}{1.5256}                           & \multicolumn{1}{r|}{0.1577}                              & \multicolumn{1}{r|}{1.0012}          & \multicolumn{1}{r|}{0.2471} & \multicolumn{1}{r|}{1.6747}                           & \multicolumn{1}{r|}{0.2341}                              & \multicolumn{1}{r|}{1.2635}          & \multicolumn{1}{r|}{0.3670}             & \multicolumn{1}{l|}{1.7875}                           & \multicolumn{1}{l|}{0.2287}                              & \multicolumn{1}{l|}{1.3449}          & \multicolumn{1}{l|}{0.3850}             & \multicolumn{1}{r|}{0.3850} \\ \hline
\end{tabular}
			\end{minipage}}
		\end{table}
		\end{widetext}

Entanglement transformation by means of local operation and classical communication (LOCC) is a fundamental toolbox, which helps to prepare the desired form of entanglement for different quantum information processing tasks. We characterised entanglement transformation for bipartite random pure states in various dimensions and explored the typical properties of catalytic transformation. We proved that a pair of randomly generated two-qutrit states is more likely to be incomparable when the difference between their entanglement content is low, whereas they become increasingly suitable for conversion as their entanglement difference grows. We also demonstrated that such a connection holds for higher dimensions as well.
The fraction of incomparable states has been shown to increase with the  increasing dimension of the subsystems, while it decreases with the increase in the number of copies of the pair of states, thereby indicating how the additional copies of the source and target states act as resources facilitating the transformation. 

For incomparable states, we resorted to the use of certain other entangled states, which act as catalysts to help further the transformation via LOCC. For a given pair of incomparable states, we Haar uniformly generated  catalytic resources, studied the entanglement properties  between their entanglement contents and the entanglement of the input-output pair, thereby establishing a universal pattern for entanglement transformation via typical states. 
We found that the average as well as minimum entanglement required to aid the transformation of an initially incomparable pair decreases with the increase in entanglement difference between the states constituting such a  pair. This result is intriguing, since the increase in entanglement difference indicates a heavier cost for the transformation to take place, which even though not possible originally, can be effected by a catalyst with lower entanglement content. Our results remain valid in all dimensions considered and also for multiple copies of the states.
Further, we reported that the entanglement content of catalysts required for  multi-copy transformations with LOCC is necessarily lower than that for the single-copy routine. However, within the state space,  such catalysts in the multi-copy domain occupy a smaller volume which implies that resources required to further multi-copy incomparable transformations are harder to generate, since they can overcome shortcomings of several copies of the original states. 

We proposed a possible classification scheme among catalyst states according to their power in the state transformation under LOCC, thereby providing a hierarchy among catalysts. Specifically, we found that there are catalysts which can make \(n+1\)-copy incomparable states comparable at the \(n\)-copy level. 
Moreover, we introduced a concept called \textit{strong catalyst} which can aid in single-copy transformations of multi-copy incomparable states. Such states prove to be most resourceful since they can make transformations possible at a single-copy level, even when multiple copies of the initial states cannot  be successful in doing so. We demonstrated the entanglement properties and distribution of such states and concluded that they exist in all dimensions. Another important feature that catalysts exhibit is the ability to effect transformations at a lower entanglement cost which we call as cost-efficient catalytic transformation. States which catalyse single-copy transformations of $2$-copy comparable states are found to possess less entanglement than the original pair of states which indicates that even though an additional copy of the states themselves can furnish the transformation, the catalyst states can actually accomplish the same task at a lower resource cost.

Notice that state transformation has been used to quantify entanglement in a unique footing (cf. \cite{Vidal_PRL_1999,Jonathan_PRA_2000}). We believe that our work contributes to such analysis in the context of typical states as to how  entanglement behaves in different dimensions and also how it can be manipulated.\\

\section{Acknowledgement}
	
	We acknowledge the support from Interdisciplinary Cyber Physical Systems (ICPS) program of the Department of Science and Technology (DST), India, Grant No.: DST/ICPS/QuST/Theme- 1/2019/23 and SM acknowledges the Ministry of Science and Technology, Taiwan (Grant No. MOST 111- 2124-M-002-013). We  acknowledge the use of \href{https://github.com/titaschanda/QIClib}{QIClib} -- a modern C++ library for general purpose quantum information processing and quantum computing (\url{https://titaschanda.github.io/QIClib}) and cluster computing facility at Harish-Chandra Research Institute. 



\bibliographystyle{apsrev4-1}
\bibliography{bib}

\end{document}